\documentclass[notitlepage]{report}

\usepackage{chngcntr}
\counterwithout{figure}{chapter}
\counterwithout{equation}{chapter}

\usepackage{titlesec}
\titleformat{\chapter}
  {\normalfont\LARGE\bfseries}{\thechapter}{1em}{}
\titlespacing*{\chapter}{0pt}{3.5ex plus 1ex minus .2ex}{2.3ex plus .2ex}

\usepackage[usenames, dvipsnames]{color}
\usepackage[dvipsnames]{xcolor}

\usepackage{geometry}
\usepackage{courier}
\usepackage{amsmath}
\usepackage{amssymb}
\usepackage{amsfonts}
\usepackage{amsthm}
\usepackage{graphicx}
\usepackage{tikz}
\usepackage{mathtools}
\usepackage[title]{appendix}
\usepackage{xcolor}
\definecolor{xLightGray}{gray}{0.9}

\newcommand{\figref}[1] {Figure \ref{fig:#1}}
\newcommand{\secref}[1] {Section \ref{sec:#1}}
\newcommand{\doublebar}[1] {\bar{\bar{#1}}}

\usepackage{caption}
\usepackage{hyperref}

\usepackage{listings}
\title{Differential Machine Learning}

\author{Brian Huge \\
brian.huge@danskebank.dk
\and Antoine Savine 
\\ antoine.savine@danskebank.dk
}
\date{Written January 2020, updated October 2020}

\begin{document}

\begingroup
\let\clearpage\relax

\maketitle



\begin{abstract}

Differential machine learning combines automatic adjoint differentiation (AAD) with modern machine learning (ML) in the context of risk management of financial Derivatives. We introduce novel algorithms for training fast, accurate pricing and risk approximations, online, in real time, with convergence guarantees. Our machinery is applicable to arbitrary Derivatives instruments or trading books, under arbitrary stochastic models of the underlying market variables. It effectively resolves computational bottlenecks of Derivatives risk reports and capital calculations.

Differential ML is a general extension of supervised learning, where ML models are trained on examples of not only inputs and labels but also \emph{differentials of labels wrt inputs}. It is also applicable in many situations outside finance, where high quality first-order derivatives wrt training inputs are available. Applications in Physics, for example, may leverage differentials known from first principles to learn function approximations more effectively. 

In finance, AAD computes \emph{pathwise differentials} with remarkable efficacy so differential ML algorithms provide extremely effective pricing and risk approximations. We can produce fast analytics in  models too complex for closed form solutions, extract the risk factors of complex transactions and trading books, and effectively compute risk management metrics like reports across a large number of scenarios, backtesting and simulation of hedge strategies, or regulations like XVA, CCR, FRTB or SIMM-MVA. 

TensorFlow implementation is available on 
\begin{center}
\url{https://github.com/differential-machine-learning}
\end{center}

\end{abstract}

\chapter*{Introduction}

Standard ML trains neural networks (NN) and other supervised ML models on punctual examples, whereas differential ML teaches them \emph{the shape} of the target function from the differentials of training labels wrt training inputs. The result is a vastly improved performance, especially in high dimension with small datasets, as we illustrate with numerical examples from both idealized and real-world contexts in \secref{Num}. 

We focus on deep learning in the main text, where the simple mathematical structure of neural networks simplifies the exposition. In the appendices, we generalize the ideas to other kind of ML models, like classic regression or principal component analysis (PCA), with equally remarkable results.
  
We posted a TensorFlow implementation on GitHub\footnote{\url{https://github.com/differential-machine-learning}}. The notebooks run on Google Colab, reproduce some of our numerical examples and discuss many practical implementation details not covered in the text. 

We could not have achieved these results without the contribution and commitment of Danske Bank's Ove Scavenius and our colleagues from Superfly Analytics, the Bank's quantitative research department. The advanced numerical results of \secref{Num} were computed with Danske Bank's production risk management system. The authors also thank Bruno Dupire, Jesper Andreasen and Leif Andersen for many insightful discussions, suggestions and comments, resulting in a considerable improvement of the contents.

\section*{Pricing approximation and machine learning}

Pricing function approximation is critical for Derivatives risk management, where the value and risk of transactions and portfolios must be computed rapidly. Exact closed-form formulas \emph{a la} Black and Scholes are only available for simple instruments and simple models. More realistic stochastic models and more complicated exotic transactions require numerical pricing by finite difference methods (FDM) or Monte-Carlo (MC), which is too slow for many practical applications. Researchers experimented with e.g. moment matching approximations for Asian and Basket options, or Taylor expansions for stochastic volatility models, as early as the 1980s. Iconic expansion results were derived in the 1990s, including Hagan's SABR formula \cite{SABR} or Musiela's swaption pricing formula in the Libor Market Model \cite{BGM}, and allowed the deployment of sophisticated models on trading desks. New results are being published regularly, either in traditional form \cite{AntonovSABR} \cite{BangSABR} or by application of advances in machine learning. 

Although pricing approximations were traditionally derived by hand, automated techniques borrowed from the fields of artificial intelligence (AI) and ML got traction in the recent years. The general format is classic supervised learning: approximate asset pricing functions $f\left(x\right)$ of a set of inputs $x$ (market variables, path-dependencies, model and instrument parameters), with a function $\hat{f} \left( x; w \right)$ subject to a collection of adjustable weights $w$, learned from a training set of $m$ examples of inputs $x^{\left(i\right)}$ (each a vector of dimension $n$) paired with labels $y^{\left(i\right)}$ (typically real numbers), by minimization of a cost function (often the mean squared error between predictions and labels). 

For example, the recent \cite{McGheeNN} and \cite{horvath2019deep} trained neural networks to price European calls\footnote{Early exploration of neural networks for pricing approximation \cite{Lo} or in the context of Longstaff-Schwartz for Bermudan and American options \cite{HaughKogan} were published over 25 years ago, although it took modern deep learning techniques to achieve the performance demonstrated in the recent works.}, respectively in the SABR model and the 'rough' volatility family of models \cite{GatheralRough}. The training sets included a vast number of examples, labeled by ground truth prices, computed by numerical methods. This approach essentially interpolates prices in parameter space. The computation of the training set takes considerable time and computational expense. The approximation is trained \emph{offline}, also at a significant computation cost, but the trained model may be reused in many different situations. Like Hagan or Musiela's expansions in their time, effective ML approximations make sophisticated models like rough Heston practically usable e.g. to simultaneously fit SP500 and VIX smiles \cite{GatVix}.

ML models generally learn approximations from training data alone, without additional knowledge of the generative simulation model or financial instrument. Although performance may be considerably improved on a case by case basis with contextual information such as the nature of the transaction, the most powerful and most widely applicable ML implementations achieve accurate approximations from data alone. Neural networks, in particular, are capable of learning accurate approximations from data, as seen in \cite{McGheeNN} and \cite{horvath2019deep} among many others. Trained NN computes prices and risks with near analytic speed. Inference is as fast as a few matrix by vector products in limited dimension, and differentiation is performed in similar time by backpropagation.

\section*{Online approximation with sampled payoffs}

While it is the risk management of large Derivatives books that initially motivated the development of pricing approximations, they also found a major application in the context of regulations like XVA, CCR, FRTB or SIMM-MVA, where the values and risk sensitivities of Derivatives trading books are repeatedly computed in many different market states. An effective pricing approximation could execute the repeated computations orders of magnitude faster and resolve the considerable bottlenecks of these computations.

However, the offline approach of \cite{McGheeNN} or \cite{horvath2019deep} is not viable in this context. Here, we learn the value of a given trading book as function of market state. The learned function is used in a set of risk reports or capital calculations and not reusable in other contexts. Such \emph{disposable} approximations are trained \emph{online}, i.e. as a part of the risk computation, and we need it performed quickly and automatically. In particular, we cannot afford the computational complexity of numerical ground truth prices.

It is much more efficient to train approximations on \emph{sampled payoffs} in place of ground truth prices, as in the classic Least Square Method (LSM) of \cite{LSM} and \cite{Carriere}. In this context, a training label is a payoff simulated on one Monte-Carlo path conditional to the corresponding input. The entire training set is simulated for a cost comparable to \emph{one} pricing by Monte-Carlo, and labels remain \emph{unbiased} (but noisy) estimates of ground truth prices (since prices are expected payoffs).

More formally, a training set of sampled payoffs consists in $m$ independent realizations $\left(x^{\left(i\right)}, y^{\left(i\right)} \right)$ of the random variables $\left(X,Y\right)$ where $X \in \mathbb{R}^n$ is the initial state and $Y \in \mathbb{R}$ is the final payoff. Informally:

\begin{align*}
\text{price} & = E\left[ {\text{payoff}\left| {\text{state}} \right.} \right]  = E\left[ {Y\left| X \right.} \right] \\
       & = {\arg {{\min }_f}E\left\{\left[ f\left( X \right) - Y \right]^2\right\}} \left( X \right)  \\
      & \approx  {\arg {{\min }_w}E\left\{\left[ \hat f\left( {X;w} \right) - Y \right]^2\right\}} \left( X \right) && {\text{ for a universal approximator } \hat f \text{, asymptotically in capacity}}\\
      & \approx {\arg \min}_w MSE\left( {{x^{\left( i \right)}},{y^{\left( i \right)}}} \right)\left( X \right) && \text{ asymptotically in the size } m \text{ of the training set} 
\end{align*}

\noindent Hence, universal approximations like neural networks, trained on datasets of sampled payoffs by minimization of the mean squared error (MSE) converge to the correct pricing function. The initial state $X$ is sampled over the domain of application for the approximation $\hat f$, whereas the final payoff $Y|X$ is sampled with a conditional MC path. See \ref{app1} for a more detailed formal exposition.

NN approximate prices more effectively than classic linear models. Neural networks are resilient in high dimension and effectively resolve the long standing \emph{curse of dimensionality} by learning regression features from data. The extension of LSM to deep learning was explored in many recent works like \cite{lapeyre2019neural}, with the evidence of a considerable improvement, in the context of Bermudan options, although the conclusions carry over to arbitrary schedules of cash-flows. We further investigate the relationship of NN to linear regression in \ref{app4}.

\section*{Training with derivatives}

We found, in agreement with recent literature, that the performance of modern deep learning remains insufficient for online application with complex transactions or trading books. A vast number of training examples (often in the hundreds of thousands or millions) is necessary to learn accurate approximations, and even a training set of sample payoffs cannot be simulated in reasonable time. Training on noisy payoffs is prone to overfitting, and unrealistic dataset sizes are necessary even in the presence of classic regularization. In addition, risk sensitivities converge considerably slower than values and often remain too approximate even with training sets in the hundreds of thousands of examples.

This article proposes to resolve these problems by training ML models on datasets \emph{augmented with differentials} of labels wrt inputs:

\[
x^{\left(i\right)} \; , \; y^{\left(i\right)} \; , \; \frac{\partial y^{\left(i\right)}}{\partial x^{\left(i\right)}}
\]

This is a somewhat natural idea, which, along with the adequate training algorithm, enables ML models to learn accurate approximations even from small datasets of noisy payoffs, making ML approximations tractable in the context of trading books and regulations. 

When learning from ground truth labels, the input $x^{\left(i\right)}$ is one example parameter set of the pricing function. If we were learning Black and Scholes' pricing function, for instance, (without using the formula, which is what we would be trying to approximate), $x^{\left(i\right)}$ would be one possible set of values for the initial spot price, volatility, strike and expiry (ignoring rates or dividends). The label $y^{\left(i\right)}$ would be the (ground thruth) call price computed with these inputs (by MC or FDM since we don't know the formula), and the derivatives labels $\partial y^{\left(i\right)} / \partial x^{\left(i\right)}$ would be the Greeks.

When learning from simulated payoffs, the input $x^{\left(i\right)}$ is one example state. In the Black and Scholes example, $x^{\left(i\right)}$ would be the spot price sampled on some present or future date $T_1 \geq 0$, called \emph{exposure date} in the context of regulations, or \emph{horizon date} in other contexts. The label $y^{\left(i\right)}$ would be the payoff of a call expiring on a later date $T_2$, sampled on that same path number $i$. The exercise is to learn a function of $S_{T_1}$ approximating the value of the $T_2$ call measured at $T_1$. In this case, the differential labels $\partial y^{\left(i\right)} / \partial x^{\left(i\right)}$ are the \emph{pathwise derivatives} of the payoff at $T_2$ wrt the state at $T_1$ on path number $i$. In Black and Scholes:

$$
    \frac{\partial y^{\left(i\right)}}{\partial x^{\left(i\right)}} = \frac{\partial \left( S_{T_2}^{\left(i\right)} - K \right)^+}{\partial S_{T_1}^{\left(i\right)}} = \frac{\partial \left( S_{T_2}^{\left(i\right)} - K \right)^+}{\partial S_{T_2}^{\left(i\right)}} \frac{\partial S_{T_2}^{\left(i\right)}}{\partial S_{T_1}^{\left(i\right)}} = 1_{\left\{ S_{T_2}^{\left(i\right)} > K \right\}} \frac{S_{T_2}^{\left(i\right)}}{S_{T_1}^{\left(i\right)}}
$$

This simple exercise exhibits some general properties of pathwise differentials. First, we computed the Black and Scholes pathwise derivative analytically with an application of the chain rule. The resulting formula is computationally efficient: the derivative is computed together with the payoff along the path, there is no need to regenerate the path, contrarily to e.g. differentiation by finite difference. This efficacy is not limited to European calls in Black and Scholes: pathwise differentials are \emph{always} efficiently computable by a systematic application of the chain rule, also known as \emph{adjoint differentiation} or AD. Furthermore, \emph{automated} implementations of AD, or AAD, perform those computations by themselves, behind the scenes.  

Secondly, ${\partial Y}/{\partial X}$ is a $T_2$ measurable random variable, and its $T_1$ expectation is $N\left(d_1\right)$, the Black and Scholes delta. This property too is general: assuming appropriate smoothing of discontinuous cash-flows, expectation and differentiation commute so risk sensitivities are expectations of pathwise differentials. Turning it upside down, pathwise differentials are unbiased (noisy) estimates of ground truth Greeks.

Therefore, we can compute pathwise differentials efficiently and use them for training as unbiased estimates of ground truth risks, irrespective of the transaction or trading book, and irrespective of the stochastic simulation model. Learning from ground truth labels is slow, but the learned function is reusable in many contexts. This is the correct manner to learn e.g. European option pricing functions in stochastic volatility models. Learning from simulated payoffs is fast, but the learned approximation is a function of the state, specific to a given financial instrument or trading book, under a given calibration of the stochastic model. This is how we can quickly approximate the value and risks of complex transactions and trading books, e.g. in the context of regulations. Differential labels vastly improve performance in both cases, as we see next. 

Classic numerical analysis applies differentials as constraints in the context of \emph{interpolation}, or penalties in the context of \emph{regularization}. Regularization generally penalises the \emph{norm} of differentials, e.g. the size of second order differentials, expressing a preference for linear functions. Our proposition is different. We do not express \emph{preferences}, we enforce differential \emph{correctness}, measured by proximity of predicted risk sensitivities to \emph{differential labels}. An application of differential labels was independently proposed in \cite{ChanMikaelWarin}, in the context of high dimensional semi-linear partial differential equations. Our algorithm is general. It applies to either ground truth learning (closely related to interpolation) or sample learning (related to regression). It consumes derivative sensitivities for ground truth learning or pathwise differentials for sample learning. It relies on an effective computation of the differential labels, achieved with automatic adjoint differentiation (AAD). 

\section*{Effective differential labels with AAD}
\label{sec:AAD}

Differential ML consumes the differential labels $\partial y^{\left(i\right)} / \partial x^{\left(i\right)}$ from an augmented training set. The differentials must be \emph{accurate} or the optimizer might get lost chasing wrong targets, and they must be computed quickly, even in high dimension, for the method to be applicable in realistic contexts. Conventional differentiation algorithms like finite differences fail on both counts. This is where the superior AAD algorithm steps in, and automatically computes the differentials of arbitrary calculations, with analytic accuracy, for a computation cost proportional to \emph{one} evaluation of the price, irrespective of dimension\footnote{This is the critical \emph{constant time} property of adjoint differentiation. It takes the time of 2 to 5 evaluations in practice to compute thousands of differentials with an efficient implementation, see \cite{savine2018modern}.}.

AAD was introduced to finance in the ground breaking 'Smoking Adjoints' \cite{Smoking}. It is closely related to backpropagation, which powers modern deep learning and has largely contributed to its recent success. In finance, AAD produces risk reports in real time, including for exotic books or XVA. In the context of Monte-Carlo or LSM, AAD produces exact pathwise differentials for a very small cost. AAD made differentials massively available in quantitative finance. Besides evident applications to instantaneous calibration or real-time risk reports, the vast amount of information contained in differentials  may be leveraged in creative ways, see e.g. \cite{AdilAAD} for an original application. 

To a large extent, differential ML is another strong application of AAD. For reasons of memory and computation efficiency, AAD always computes differentials path by path when applied with Monte-Carlo, effectively estimating risk sensitivities in a vast number of different scenarios. Besides its formidable speed and accuracy, AAD therefore produces a massive amount of information. Risk reports correspond to \emph{average} sensitivities across paths, they only provide a much \emph{flattened} view of the pathwise differential information. Differential ML, on the other hand, leverages its full extent in order to learn value and risk, not as fixed numbers only relevant in the current state, but as \emph{functions} of state capable of computing prices and Greeks very quickly in different market scenarios.

In the interest of brevity, we refer to \cite{savine2018modern} for a comprehensive description of AAD, including all details of how training differentials were obtained in this study, or the video tutorial \cite{AADVideo}, which explains its main ideas in 15 minutes.        

The main article is voluntarily kept rather concise. Practical implementation details are deferred to the online notebook, and mathematical formalism is treated in the appendices along with generalizations and extensions. We present differential ML in \secref{DiffMachineLearning} in the context of feedforward neural networks, numerical results in \secref{Num} and important extensions in \secref{multi}. \ref{app1} deploys the mathematical formalism of the machinery. \ref{app2} introduces differential PCA and \ref{app3} applies differential ML as a superior regularization in the context of classic linear regression. \ref{app4} discusses neural architectures and asymptotic control algorithms with convergence guarantees necessary for online operation.

\chapter{Differential Machine Learning}
\label{sec:DiffMachineLearning}

This section describes differential training in the context of feedforward neural networks, although everything carries over to NN of arbitrary complexity in a straightforward manner. At this stage, we assume the availability of a training set augmented with differential labels. The dataset consists of arbitrary schedules of cash-flows simulated in an arbitrary stochastic model. Because we learn from simulated data alone, there are no restrictions on the sophistication of the model or the complexity of the cash-flows. The cash-flows of the transaction or trading book could be described with a general scripting language, and the model could be a hybrid 'model of everything' often used for e.g. XVA computations, with dynamic parameters calibrated to current market data.

The text focuses on a mathematical and qualitative description of the algortihm, leaving the discussion of practical implementation to the online notebook\footnote{\url{https://github.com/differential-machine-learning/notebooks/blob/master/DifferentialML.ipynb}}, along with TensorFlow implementation code.

\section{Notations}

\subsection{Feedforward equations}

Let us first introduce notations for the description of feedforward networks. Define the input (row) vector $x\in \mathbb R^{n}$ and the predicted value $y\in\mathbb R$. For every layer $l=1,\ldots,L$ in the network, define a scalar 'activation' function $g_{l-1}:\mathbb R\to \mathbb R$. Popular choices are relu, elu and softplus, with the convention $g_0(x)=x$ is the identity. The notation $g_{l-1}(x)$ denotes elementwise application. We denote $w_l\in\mathbb R^{n_{l-1}\times n_{l}},b_l \in \mathbb R^{n_{l}}$ the weights and biases of layer $l$. 

The network is defined by its feedforward equations:

\begin{eqnarray}
\nonumber
z_0 & = & x \\
\label{eq:ffn}
z_{l} & = & g_{l-1}\left(z_{l-1}\right) w_{l} + b_{l}\quad , l=1,\ldots,L \\
\nonumber
y & = & z_{L}
\end{eqnarray}

\noindent where $z_l \in \mathbb R^{n_l}$ is the row vector containing the $n_l$ pre-activation values, also called \emph{units} or \emph{neurons}, in layer $l$. \figref{ffn} illustrates a feedforward network with $L=3$ and $n=n_0=3, n_1=5, n_2=3, n_3=1$, together with backpropagation.

\subsection{Backpropagation}

Feedforward networks are efficiently differentiated by backpropagation, which is generally applied to compute the derivatives of some some cost function wrt the weights and biases for optimization. For now, we are not interested in those differentials, but in the differentials of the \emph{predicted} value $y=z_L$ wrt the \emph{inputs} $x=z_0$. Recall that inputs are states and predictions are prices, hence, these differentials are predicted risk sensitivities (\emph{Greeks}), obtained by differentiation of the lines in \eqref{eq:ffn}, in the reverse order:

\begin{eqnarray}
\nonumber
\bar z_{L} & = & \bar y = 1\\
\label{eq:back} 
\bar z_{l-1} & = & \left(\bar z_{l}w_l^T\right)\circ g_{l-1}^{\prime}\left(z_{l-1}\right)\quad , l=L,\ldots,1 \\
\nonumber
\bar x & = & \bar z_0
\end{eqnarray}

\noindent with the \emph{adjoint} notation $\bar x = \partial y / \partial x,\bar z_l = \partial y / \partial z_l, \bar y = \partial y / \partial y=1$ and $\circ$ is the elementwise (Hadamard) product. 

\begin{figure}[htp]
\centering
\includegraphics[scale=0.6]{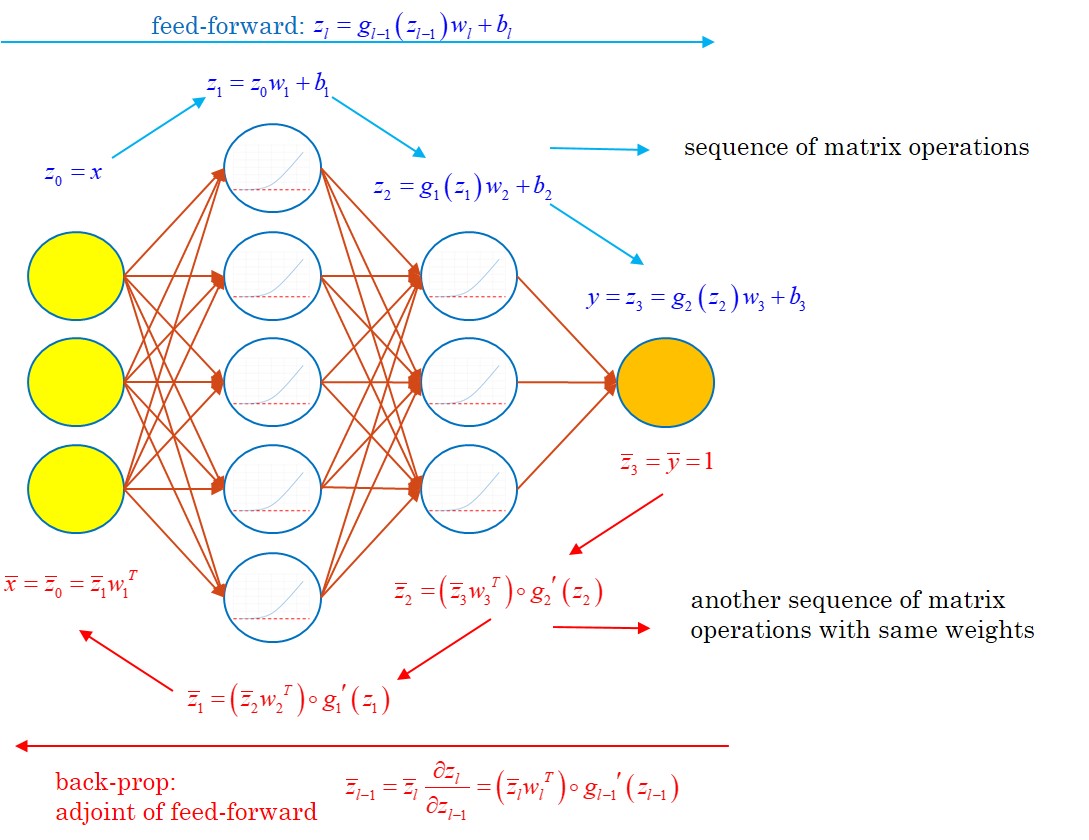}
\caption{feedforward neural network with backpropagation}
\label{fig:ffn}
\end{figure}

Notice, the similarity between \eqref{eq:ffn} and  \eqref{eq:back}. In fact, backpropagation defines a second feedforward network with inputs $\bar y, z_0,\ldots, z_L$ and output $\bar x\in\mathbb R^{n}$, where the weights are shared with the first network and the units in the second network are the adjoints of the corresponding units in the original network.

Backpropagation is easily generalized to arbitrary network architectures, as explained in deep learning literature. Generalized to arbitrary computations unrelated to deep learning or AI, backpropagation becomes AD, or AAD when implemented automatically\footnote{See video tutorial \cite{AADVideo}.}. Modern frameworks like TensorFlow include an implementation of backpropagation/AAD and implicitly invoke it in training loops.

\section{Twin networks}
\label{sec:Shape}

We can combine feedforward \eqref{eq:ffn} and backpropagation \eqref{eq:back} equations into a single network representation, or \emph{twin network}, corresponding to the computation of a prediction (approximate price) together with its differentials wrt inputs (approximate risk sensitivities).

The first half of the twin network (\figref{twinNN}) is the original network, traversed with feedforward induction to predict a value. The second half is computed with the backpropagation equations to predict risk sensitivities. It is the mirror image of the first half, with shared connection weights. 

A mathematical description of the twin network is simply obtained by concatenation of equations \eqref{eq:ffn} and \eqref{eq:back}. The evaluation of the twin network returns a predicted value $y$, and its differentials $\bar x$ wrt the $n_0=n$ inputs $x$. The combined computation evaluates a feedforward network of twice the initial depth. Like feedforward induction, backpropagation computes a sequence of matrix by vector products. The twin network, therefore, predicts prices and risk sensitivities for twice the computation complexity of value prediction alone, irrespective of the number of risks. Hence, a trained twin net approximates prices and risk sensitivities, wrt potentially many states, in a particularly efficient manner.  
Note from \eqref{eq:back} that the units of the second half are activated with the differentials $g_l'$ of the original activations $g_l$. If we are going to backpropagate through the twin network, we need continuous activation throughout. Hence, the initial activation must be $C^1$, ruling out, e.g. ReLU.

\begin{figure}[htp]
\centering
\includegraphics[scale=0.60]{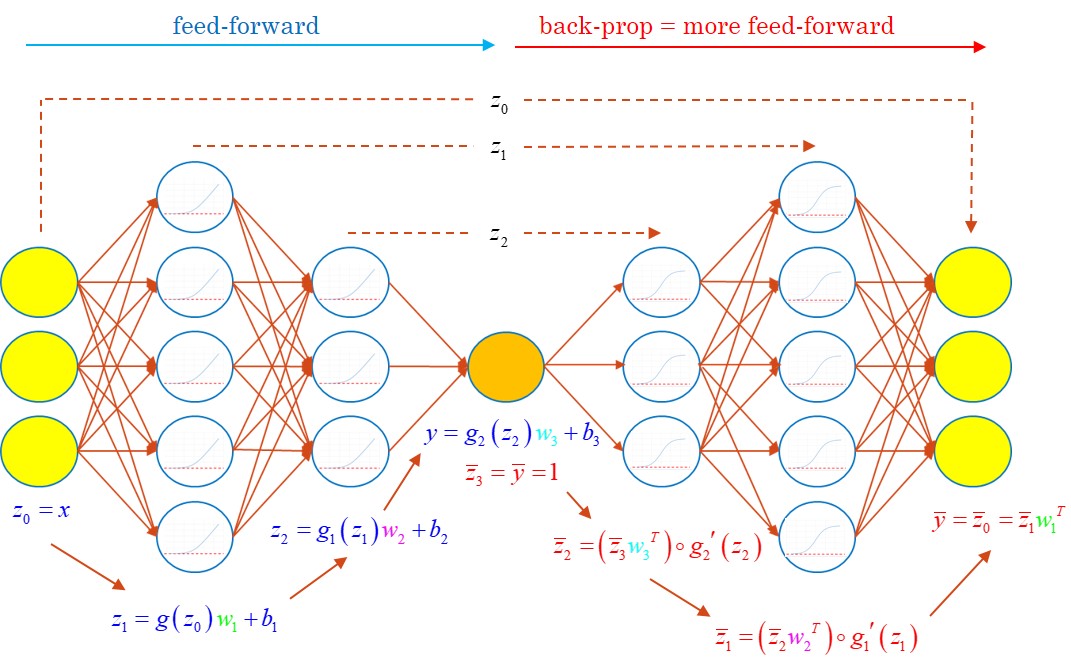}
\caption{twin network}
\label{fig:twinNN}
\end{figure}

\subsection{Training with differential labels}
\label{sec:train}

The purpose the twin network is to estimate the correct pricing function $f \left( x \right)$ by an approximate function $\hat f\left(x;\left\{ w_l, b_l\right\}_{l=1,\ldots,L}\right)$. It learns optimal weights and biases from an augmented training set $\left( x^{\left( i \right)}, y^{\left( i \right)}, \bar x^{\left( i \right)} \right)$, where $ \bar x^{\left( i \right)} = \partial y^{\left( i \right)} / \partial x^{\left( i \right)}$ are the differential labels. 

Here, we describe the mechanics of differential training and discuss its effectiveness. As is customary with ML, we stack training data in matrices, with examples in rows and units in columns:
\[
X = \left[
\begin{array}{c}
x^{(1)}\\
\vdots\\
x^{(m)}
\end{array}
\right]
\in \mathbb R^{m\times n}\quad
Y = \left[
\begin{array}{c}
y^{(1)}\\
\vdots\\
y^{(m)}
\end{array}
\right]
\in \mathbb R^{m}\quad
\bar X = \left[
\begin{array}{c}
\bar x^{(1)}\\
\vdots\\
\bar x^{(m)}
\end{array}
\right]
\in \mathbb R^{m\times n}
\]
Notice, the equations \eqref{eq:ffn} and \eqref{eq:back} identically apply to matrices or row vectors. Hence, the evaluation of the twin network computes the matrices:
\[
Z_l = \left[
\begin{array}{c}
z^{(1)}_l\\
\vdots\\
z^{(m)}_l
\end{array}
\right]
\in \mathbb R^{m\times n_l}\quad \text{and} \quad
\bar Z_l = \left[
\begin{array}{c}
\bar z^{(1)}_l\\
\vdots\\
\bar z^{(m)}_l
\end{array}
\right]
\in \mathbb R^{m\times n_l}
\]

\noindent respectively in the first and second half of its structure. Training consists in finding weights and biases minimizing some cost function $C$: $ \left\{w_l,b_l\right\}_{l=1,\ldots,L} = \arg \min C\left(\left\{w_l,b_l\right\}_{l=1,\ldots,L}\right) $.

\subsubsection{Classic training with payoffs alone}

Let us first recall classic deep learning. We have seen that the approximation obtained by global minimization of the MSE converges to the correct pricing function (modulo finite capacity bias), hence:

$$
    C\left(\left\{w_l,b_l\right\}_{l=1,\ldots,L}\right)=MSE= \frac 1 m\left(Z_L-Y\right)^T\left(Z_L-Y\right)
$$

The second half of the twin network does not affect cost, hence, training is performed by backpropagation through the standard feedforward network alone. The many practical details of the optimization are covered in the online notebook. 

\subsubsection{Differential training with differentials alone}

Let us change gears and train with pathwise differentials $\bar x ^{\left( i \right)}$ instead of payoffs $y ^{\left( i \right)}$, by minimization of the MSE (denoted $\overline{MSE}$) between the differential labels (pathwise differentials) and predicted differentials (estimated risk sensitivities):

$$
    C\left(\left\{w_l,b_l\right\}_{l=1,\ldots,L}\right)=\overline{MSE}= \frac 1 {m} \textrm{tr}\left[ \left(\bar Z_0-\bar X\right)^T\left(\bar Z_0-\bar X\right) \right] 
$$

Here, we must evaluate the twin network in full to compute $\bar Z_0$, effectively doubling the cost of training. Gradient-based methods minimize $\overline{MSE}$ by backpropagation through the twin network, effectively accumulating second-order differentials in its second half. A deep learning framework, like TensorFlow, performs this computation seamlessly. As we have seen, the second half of the twin network may represent backpropagation, in the end, this is just another sequence of matrix operations, easily differentiated by another round of backpropagation, carried out silently, behind the scenes. The implementation in the demonstration notebook is identical to training with payoffs, safe for the definition of the cost function. TensorFlow automatically invokes the necessary operations, evaluating the feedforward network when minimizing $MSE$ and the twin network when minimizing $\overline {MSE}$. 

In practice, we must also assign appropriate weights to the costs of wrong differentials in the definition of the $\overline {MSE}$. This is discussed in the implementation notebook, and in more detail in \ref{app2}. 

Let us now discuss what it \emph{means} to train approximations by minimization of the $\overline{MSE}$ between pathwise differentials $\bar x^{\left(i\right)} = \partial y^{\left(i\right)}  / \partial x^{\left(i\right)} $ and predicted risks $ \partial \hat f \left( x^{\left(i\right)} \right) /  \partial x^{\left(i\right)}$. Given appropriate smoothing\footnote{
Pathwise differentials of discontinuous payoffs like digitals or barriers are not well defined, and it follows that the risk sensitivities of these instruments cannot be reliably computed with Monte-Carlo, with AAD or otherwise. This is a well-known problem in the industry, generally resolved by \emph{smoothing}, i.e. the approximation of discontinuous cash-flows with close continuous ones, like tight call spreads in place of digitals or soft barriers in place of hard barriers. Smoothing is a common practice among option traders, and it is related to \emph{fuzzy logic}, as demonstrated in \cite{SavineFuzzyLogic}, which also presents the theoretical and practical details of smoothing methodologies, and proposes a systematic smoothing algorithm based on fuzzy logic.
}, expectation and differentiation commute so the (true) risk sensitivities are expectations of pathwise differentials:

$$
    \frac{\partial f \left( x \right)} { \partial x} = \frac{\partial E \left[ Y | X=x \right]} { \partial x} = E \left[ \frac{\partial Y} { \partial X} | X=x \right]
$$

It follows that pathwise differentials are unbiased estimates of risk sensitivities, and approximations trained by minimization of the $\overline{MSE}$ converge (modulo finite capacity bias)  to a function with correct differentials, hence, the right pricing function, modulo an additive constant.

Therefore, we can choose to train by minimization of value or derivative errors, and converge near the correct pricing function all the same. This consideration is, however, an asymptotic one. Training with differentials converges near the same approximation, but it converges much faster, allowing us to train accurate approximations with much smaller datasets, as we see in the numerical examples, because:

\begin{description}

\item [The effective size of the dataset is much larger] evidently, with $m$ training examples we have $n m$ differentials ($n$ being the dimension of the inputs $x^{\left(i\right)}$ ). With AAD, we effectively simulate a much larger dataset for a minimal additional cost, especially in high dimension (where classical training struggles most).

\item [The neural nets picks up the \emph{shape} of the pricing function] learning from slopes rather than points, resulting in much more stable and potent learning, even with few examples.

\item [The neural approximation learns to produce correct Greeks] by construction, not only correct values. By learning the correct shape, the ML approximation also correctly orders values in different scenarios, which is critical in applications like value at risk (VAR) or expected loss (EL), including for FRTB. 

\item [Differentials act as an effective, bias-free regularization] as we see next.

\end{description}

\subsubsection{Differential training with everything}

The best numerical results are obtained in practice by combining values and derivatives errors in the cost function:

$$
    C= MSE + \lambda \overline{MSE}
$$

\noindent which is the one implemented in the demonstration notebook, with the two previous strategies as particular cases. Notice, the similarity with classic regularization of the form $C=MSE + \lambda \text { } penalty$. Ridge (Tikhonov) and Lasso regularizations impose a penalty for large weights (respectively in $L^2$ and $L^1$ metrics), effectively preventing overfitting small datasets by stopping attempts to fit noisy labels. In return, classic regularization reduces the effective capacity of the model and introduces a bias, along with a strong dependency on the hyperparameter $\lambda$. This hyperparameter controls regularization strength and tunes the vastly documented bias-variance tradeoff. If one sets $\lambda$ too high, their trained approximation ends up a horizontal line.

Differential training also stops attempts to fit noisy labels, with a penalty for wrong differentials. It is, therefore, a form of regularization, but a very different kind. It doesn't introduce bias, since we have seen that training on differentials \emph{alone} converges to the correct approximation too. This breed of regularization comes without bias-variance tradeoff. It reduces variance for free. Increasing $\lambda$ hardly affects results in practice. 

Differential regularization is more similar to \emph{data augmentation} in computer vision, which is, in turn, a more powerful regularization. Differentials are additional training data. Like data augmentation, differential regularization reduces variance by increasing the size of the dataset for little cost. Differentials are new data of a different kind, and it shares inputs with existing data, but it reduces variance all the same, without introducing bias.

\chapter{Numerical results}
\label{sec:Num}
Let us now review some numerical results and compare the performance of differential and conventional ML. We picked three examples from relevant textbook and real-world situations, where neural networks learn pricing and risk approximations from small datasets.

We kept neural architecture constant in all the examples, with four hidden layers of 20 softplus-activated units. We train neural networks on mini-batches of normalized data, with the ADAM optimizer and a one-cycle learning rate schedule. The demonstration notebook and appendices discuss all the details. A differential training set takes 2-5 times longer to simulate with AAD, and it takes twice longer to train twin nets than standard ones. In return, we are going to see that differential ML performs up to thousandfold better on small datasets.

\section{Basket options}

The first (textbook) example is a basket option in a correlated Bachelier model for seven assets\footnote{This example is reproducible on the demonstration notebook, where the number of assets is configurable, and the covariance matrix and basket weights are re-generated randomly.}:
\[
dS_t = \sigma \, dW_t
\]
where $S_t\in \mathbb R^7$ and $dW_t^jdW_t^k= \rho_{jk}$. The task is to learn the pricing function of a 1y call option on a basket, with strike 110 (we normalized asset prices at 100 without loss of generality and basket weights sum to 1). The basket price is also Gaussian in this model; hence, Bachelier's formula gives the correct price. This example is also of particular interest because, although the input space is seven-dimensional, we know from maths that actual pricing is one-dimensional. Can the network learn this property from data?

\begin{figure}[htp]
    \centerline{\includegraphics[scale=0.4]{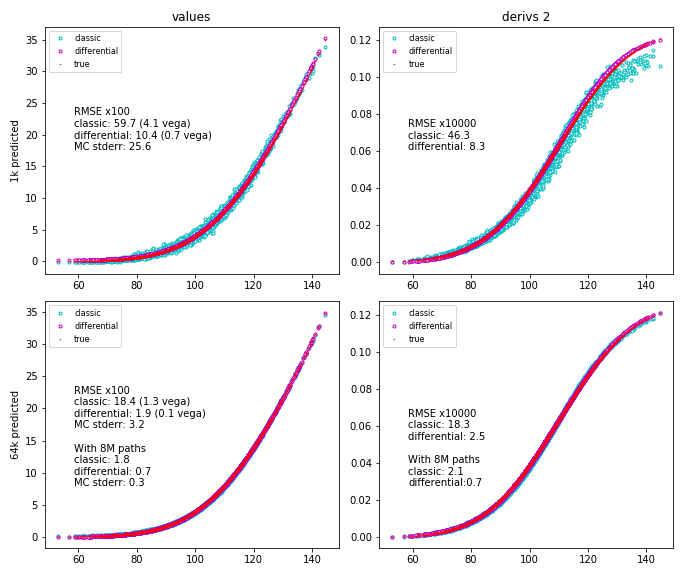}}
    \caption{basket option in Bachelier model, dimension 7}
    \label{fig:bachelier}
\end{figure}

We have trained neural networks and predicted values and derivatives in 1024 independent test scenarios, with initial basket values on the horizontal axis and option prices/deltas on the vertical axis (we show one of the seven derivatives), compared with the correct results computed with Bachelier's formula. We trained networks on 1024 (1k) and 65536 (64k) paths, with cross-validation and early stopping. The twin network with 1k examples performs better than the classical net with 64k examples for values, and a lot better for derivatives. In particular, it learned that the option price and deltas are a fixed function of the basket, as evidenced by the thinness of the approximation curve. The classical network doesn't learn this property well, even with 64k examples. It overfits training data and predicts different values or deltas for various scenarios on the seven assets with virtually identical baskets.

We also compared test errors with standard MC errors (also with 1k and 64k paths). The main point of pricing approximation is to avoid nested simulations with similar accuracy. We see that the error of the twin network is, indeed, close to MC. Classical deep learning error is an order of magnitude larger. Finally, we trained with eight \emph{million} samples, and verified that both networks converge to similarly low errors (\emph{not} zero, due to finite capacity) while MC error converges to zero. The twin network gets there hundreds of times faster.

All those results are reproduced in the online TensorFlow notebook.

\section{Worst-of autocallables}

As a second (real-world) example, we approximate an exotic instrument, a four-underlying version of the popular worst-of autocallable trade, in a more complicated model, a collection of 4 correlated local volatility models \emph{a la} Dupire: 

\[
dS_t^j = \sigma_j\left(t,S_t^j\right)\, dW_t^j \quad j=1,\ldots,4
\]

\noindent where $dW_t^jdW_t^k= \rho_{jk}$. The example is relevant, not only due to popularity, but also, because of the stress path-dependence, barriers and massive final digitals impose on numerical models. Appropriate smoothing was applied so pathwise differentials are well defined.

We do not have a closed form solution for reference, so performance is measured against nested Monte-Carlo simulations (a \emph{very} slow process). In \figref{worstof}, we show prediction results for 128 independent examples, with correct numbers on the horizontal axis, as given by the nested simulations, and predicted results on the vertical axis. Performance is measured by distance to the 45deg line. 

The classical network is trained on 32768 (32k) samples, without derivatives, with cross-validation and early stopping. The twin network is trained on 8192 (8k) samples with pathwise derivatives produced with AAD. Both sets were generated in around 0.4 sec in Superfly, Danske Bank's proprietary derivatives pricing and risk management system. 

\begin{figure}[htp]
    \centering
    \includegraphics[scale=0.7]{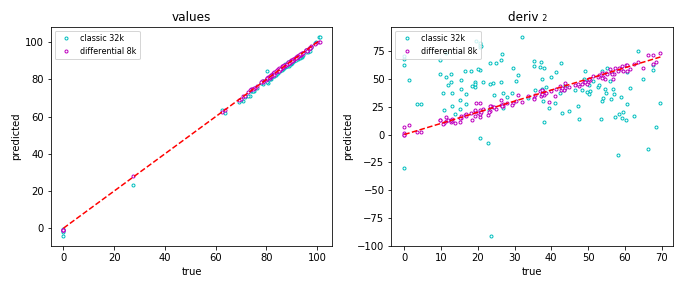}
    \includegraphics[scale=0.7]{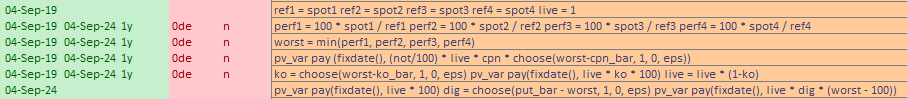}
    
    \caption{worst-of-four autocallable with correlated local volatility models}
    \label{fig:worstof}
\end{figure}

\figref{worstof} shows the results for the value and the delta to the second underlying, together with the script for the instrument, written in Danske Bank's Jive scripting language. Note that the barriers and the digitals are explicitly smoothed with the keyword 'choose'. It is evident that the twin network with only 8k training data produces a virtually perfect approximation in values, and a decent approximation on deltas. The classical network also approximates values correctly, although not on a straight line, which may cause problems when ordering is critical, e.g. for expected loss or FRTB. Its deltas are essentially random, which rules them out for approximation of risk sensitivities, e.g. for SIMM-MVA. 

Absolute standard errors are 1.2 value and 32.5 delta with the classical network with 32k examples, respectively 0.4 and 2.5 with the differential network trained on 8k examples. For comparison, the Monte-Carlo pricing error is 0.2 with 8k paths, similar to the twin net. The error on the classical net, with 4 times the training size, is larger for values and order of magnitude larger for differentials.

\section{Derivatives trading books}

For the last example, we picked a real netting set from Danske Bank's portfolio, including single and cross currency swaps and swaptions in 10 different currencies, eligible for XVA, CCR or other regulated computations. Simulations are performed in Danske Bank's model of everything (the 'Beast'), where interest rates are simulated each with a four-factor version of Andreasen's take on multi-factor Cheyette \cite{BackToTheFuture}, and correlated between one another and with forex rates.

This is an important example, because it is representative of how we want to apply twin nets in the real world. In addition, this is a stress test for neural networks. The Markov dimension of the four-factor non-Gaussian Cheyette model is 16 per currency, that is 160 inputs, 169 with forexes, and over 1000 with all the path-dependencies in this real-world book. Of course, the value effectively only depends on a small number of combinations of inputs, something the neural net is supposed to identify. In reality, the extraction of effective risk factors is considerably more effective in the presence of differential labels (see \ref{app2}), which explains the results in \figref{nettingset}. 

\figref{nettingset} shows the values predicted by a twin network trained on 8192 (8k) samples with AAD pathwise derivatives, compared to a vanilla net, trained on 65536 (64k) samples, all simulated in Danske Bank's XVA system.  The difference in performance is evident in the chart. The twin approximation is virtually perfect with on only 8k examples. The classical deep approximation is much more rough with 64k examples. As with the previous example, the predicted values for an independent set of 128 examples are shown on the vertical axis, with correct values on the horizontal axis. The 'correct' values for the chart were produced with nested Monte-Carlo overnight. The entire training process for the twin network (on entry level GPU), including the generation of the 8192 examples (on multithreaded CPU), took a few seconds on a standard workstation.

\begin{figure}[htp]
    \centering
     \includegraphics[scale=0.6]{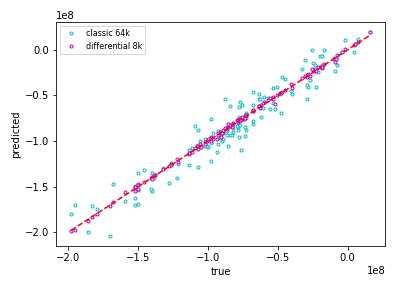}    
    \caption{real-world netting set -- twin network trained on 8k samples vs classical net trained on 64k samples}
    \label{fig:nettingset}
\end{figure}

We have shown in this figure the predicted values, not derivatives, because we have too many of them, often wrt obscure model parameters like accumulated covariances in Cheyette. For these derivatives to make sense, they must be turned into market risks by application of inverse Jacobian matrices \cite{savineJacobian}, something we skipped in this exercise.

Standard errors are 12.85M with classical 64k and 1.77M with differential 8k, for a range of 200M for the 128 test examples, generated with the calibrated hybrid model. On this example too, twin 8k error is very similar to the Monte-Carlo pricing error (1.70M with 8k paths). Again in this very representative example, the twin network has the same degree of approximation as orders of magnitude slower nested Monte-Carlo.

\chapter{Extensions}
\label{sec:multi}
We have presented algorithms in the context of single value prediction to avoid confusion and heavy notations. To conclude, we discuss two advanced extensions, allowing the network to predict multiple values and higher-order derivatives simultaneously.

\section{Multiple outputs}
One innovation in \cite{horvath2019deep} is to predict call prices of multiple strikes and expiries in a single network, exploiting correlation and shared factors, and encouraging the network to learn global features like no-arbitrage conditions. We can combine our approach with this idea by an extension of the twin network to compute multiple predictions, meaning $n_L>1$ and $y=z_L\in\mathbb R^{n_{L}}$. The adjoints are no longer well defined as vectors. Instead, we now define them as directional differentials wrt some specified linear combination of the outputs $y c^T$ where $c\in \mathbb R^{n_L}$ has the coordinates of the desired direction in a row vector:

$$\bar x = \frac{\partial{y c^T}}{\partial x},\bar z_l = \frac{\partial{y c^T}}{\partial z_l},\bar y = \frac{\partial{y c^T}}{\partial y}=c$$

Given a direction $c$, all the previous equations apply identically, except that the boundary condition for $\bar y$ in the backpropagation equations is no longer the number 1, but the row vector $c$. For example, $c = e_1$ means that adjoints are defined as derivatives of the first output $y_1$. We can repeat this for $c=e_1,\ldots,e_n$ to compute the derivatives of all the outputs wrt all the inputs $\partial y / \partial x \in \mathbb R^{n_L\times n}$, i.e the Jacobian matrix. Written in matrix terms, the boundary is the identity matrix $I\in \mathbb R^{n_L\times n_L}$ and the backpropagation equations are written as follows:
\begin{eqnarray}
\nonumber
\bar z_{L} & = & \bar y = I\\
\nonumber
\bar z_{l-1} & = & \left(\bar z_{l}w_l^T\right)\circ g_{l-1}^{\prime}\left(z_{l-1}\right),\quad l=L,\ldots,1 \\
\nonumber
\bar x & = & \bar z_0
\end{eqnarray}
where $\bar z_l\in\mathbb R^{n_{L}\times n_l}$ . In particular, $\bar x\in\mathbb R^{n_{L}\times n}$ is (indeed) the Jacobian matrix $\partial y / \partial x$. To compute a full Jacobian, the theoretical order of calculations is $n_L$ times the vanilla network. Notice however, that the implementation of the multiple backpropagation in the matrix form above on a system like TensorFlow automatically benefits from CPU or GPU parallelism. Therefore, the additional computation complexity will be experienced as sublinear.

\section{Higher order derivatives}
The twin network can also predict higher-order derivatives. For simplicity, revert to the single prediction case where $n_L=1$. The twin network predicts $\bar x$ as a function of the input $x$. The neural network, however, doesn't know anything about derivatives. It just computes numbers by a sequence of equations. Hence, we might as well consider the prediction of differentials as multiple outputs. 

As previously, in what is now considered a multiple prediction network, we can compute the adjoints of the outputs $\bar x$ in the twin network. These are now \emph{the adjoints of the adjoints}:
\[
\doublebar{x} \equiv \frac {\partial \bar x c^T}{\partial x}\in \mathbb R^n
\]
\noindent in other terms, the Hessian matrix of the value prediction $y$. Note that the original activation functions must be $C^2$ for this computation. The computation of the full Hessian is of order $n$ times the original network. These additional calculations generate a lot more data, one value, $n$ derivatives and $\frac 12 n \left(n+1\right)$ second-order derivatives for the cost of $2n$ times the value prediction alone. In a parallel system like TensorFlow, the experience also remains sublinear. We can extend this argument to arbitrary order $q$, with the only restriction that the (original) activation functions are $C^q$. 

\chapter*{Conclusion}
\label{sec:Conclusion}

Throughout our analysis we have seen that 'learning the correct shape' from differentials is crucial to the performance of regression models, including neural networks, in such complex computational tasks as the pricing and risk approximation of arbitrary Derivatives trading books. The \emph{unreasonable effectiveness} of what we called 'differential machine learning' permits to accurately train ML models on a small number of simulated payoffs, in realtime, suitable for online learning. Differential networks apply to real-world problems, including regulations and risk reports with multiple scenarios. Twin networks predict prices and Greeks with almost analytic speed, and their empirical test error remains of comparable magnitude to nested Monte-Carlo. 

Our machinery learns from data alone and applies in very general situations, with arbitrary schedules of cash-flows, scripted or not, and arbitrary simulation models. Differential ML also applies to many families of approximations, including classic linear combinations of fixed basis functions, and neural networks of arbitrary complex architecture. Differential training consumes differentials of labels wrt inputs and requires clients to somehow provide high-quality first-order differentials.  In finance, they are obtained with AAD, in the same way we compute Monte-Carlo risk reports, with analytic accuracy and very little computation cost.

One of the main benefits of twin networks is their ability to learn effectively from small datasets. Differentials inject meaningful additional information, eventually resulting in \emph{better} results with small datasets of 1k to 8k examples than can be obtained otherwise with training sets orders of magnitude larger. Learning effectively from small datasets is critical in the context of e.g. regulations, where the pricing approximation must be learned quickly, and the expense of a large training set cannot be afforded.

The penalty enforced for wrong differentials in the cost function also acts as a very effective regularizer, superior to classical forms of regularization like Ridge, Lasso or Dropout, which enforce arbitrary penalties to mitigate overfitting, whereas differentials meaningfully augment data. Standard regularizers are very sensitive to the regularization strength  $\lambda$, a manually tweaked hyperparameter. Differential training is virtually insensitive to $\lambda$, because, even with infinite regularization, we train on derivatives alone and still converge to the correct approximation, modulo an additive constant. 

\ref{app2} and \ref{app3} apply the same ideas to respectively PCA and classic regression. In the context of regression, differentials act as a very effective regularizer. Like Tikhonov regularization, differential regularization is analytic and works SVD. \ref{app3} derives a variation of the normal equation adjusted for differential regularization. Unlike Tikhonov, differential regularization does not introduce bias. Differential PCA, unlike classic PCA, is able to extract from data the principal risk factors of a given transaction, and it can be applied as a preprocessing step to safely reduce dimension without loss of relevant information.

Differential training also appears to stabilize the training of neural networks, and improved resilience to hyperparameters like network architecture, seeding of weights or learning rate schedule was consistently observed, although to explain exactly why is a topic for further research.

Standard machine learning may often be considerably improved with contextual information not contained in data, such as the nature of the relevant features from knowledge of the transaction and the simulation model. For example, we know that the continuation value of a Bermudan option on some call date mainly depends on the swap rate to maturity and the discount rate to the next call. We can learn pricing functions much more effectively with hand engineered features. But it has to be done manually, on a case by case basis, depending on the transaction and the simulation model. If the Bermudan model is upgraded with stochastic volatility, volatility state becomes an additional feature that cannot be ignored, and hand-engineered features must be updated. Differential machine learning learns just as well, or better, from data alone, the vast amount of information contained in pathwise differentials playing a role similar, and sometimes more effectively, to manual adjustments from contextual information. 

Differential machine learning is similar to data augmentation in computer vision, a technique consistently applied in that field with documented success, where multiple labeled images are produced from a single one, by cropping, zooming, rotation or recoloring. In addition to extending the training set for a negligible cost, data augmentation encourages the ML model to learn important invariances. Similarly, derivatives labels, not only increase the amount of information in the training set, but also encourage the model to learn the \emph{shape} of the pricing function. 
    
\pagebreak 

\part*{Appendices}

\appendix
\renewcommand{\thechapter}{Appx \arabic{chapter}}
\counterwithout{section}{chapter}
\setcounter{section}{0}
\chapter{Learning Prices from Samples}
\label{app1}
\section*{Introduction}

When learning Derivatives pricing and risk approximations, the main computation load belongs to the simulation of the training set. For complex transactions and trading books, it is not viable to learn from examples of ground truth prices. True prices are computed numerically, generally by Monte-Carlo. Even a small dataset of say, 1000 examples, is therefore simulated for the computation cost of 1000 Monte-Carlo pricings, a highly unrealistic cost in a practical context. Alternatively, \emph{sample} datasets \emph{a la} Longstaff-Schwartz (2001) are produced for the computation cost of \emph{one} Monte-Carlo pricing, where each example is not a ground truth price, but one sample of the payoff, simulated for the cost of one Monte-Carlo path. This methodology, also called LSM (for Least Square Method as it is called in the founding paper) simulates training sets in realistic time and allows to learn pricing approximations in realistic time.

This being said, we now expect the machine learning model to learn correct pricing functions without having ever seen a price. Consider a simple example: to learn the pricing function for a European call in Black and Scholes, we simulate a training set of call payoffs $Y^{\left( i \right)} = \left(S_{T_2}^{(i)} - K \right)^+$ given initial states $X^{\left( i \right)} = S_{T_1}^{(i)}$. The result is a random looking cloud of points $X^{\left( i \right)}, Y^{\left( i \right)}$, and we expect the machine to learn from this data the correct pricing function given by Black and Scholes' formula.

\begin{figure}[htp]
\centering
\includegraphics[scale=0.5]{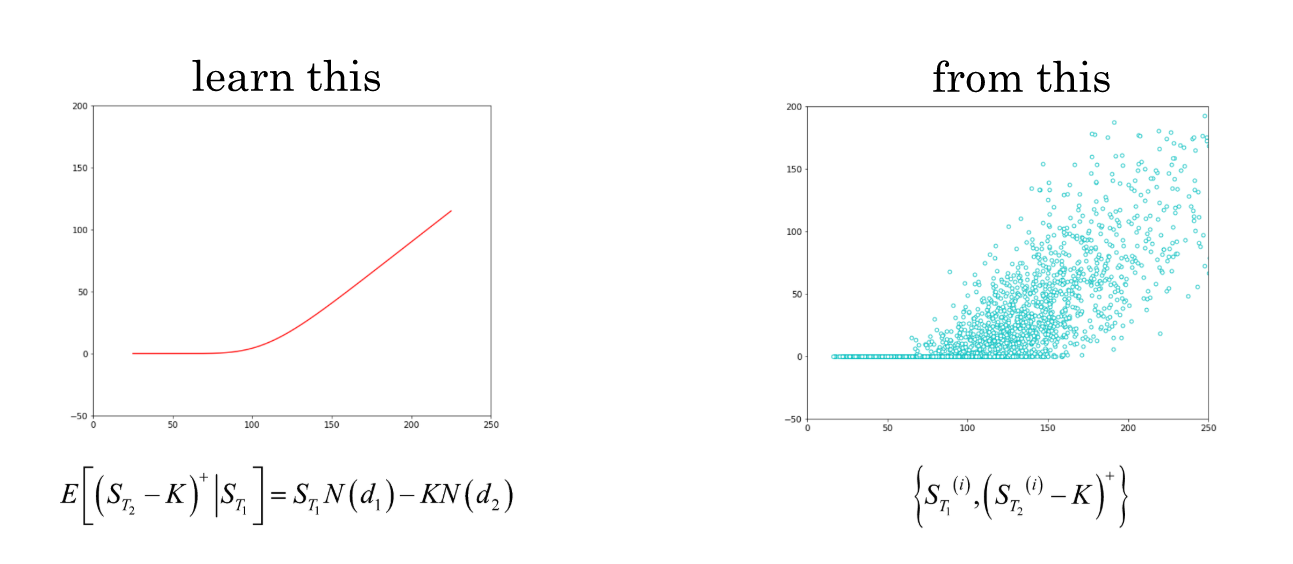}
\end{figure}

It is not given at all, and it may even seem somewhat magical, that training a machine learning model on this data should converge to the correct function. When we train on ground true prices, we essentially interpolate prices in input space, where it is clear and intuitive that arbitrary functions are approximated to arbitrary accuracy by growing the size of the training set and the capacity of the model. In fact, the same holds with LSM datasets, and this appendix discusses some important intuitions and presents sketches of mathematical proof of why this is the case.

In the first section, we recall LSM in detail\footnote{Omitting the recursive part of the algorithm, specific to early exerciseable Derivatives.} and frame it in machine learning terms. Readers familiar with the Longstaff-Schwartz algorithm may browse through this section quickly, although skipping it altogether is not recommended, this is where we set important notations. In the second section, we discuss universal approximators, formalize their training process on LSM samples, and demonstrate convergence to true prices. In the third section, we define pathwise differentials, formalize differential training and show that it too converges to true risk sensitivities. 

The purpose of this document is to explain and formalize important mathematical intuitions, not to provide complete formal proofs. We often skip important mathematical technicalities so our demonstrations should really be qualified as 'sketches of proof'.

\section{LSM datasets}

    \subsection{Markov States}

        \subsubsection{Model state}

        First, we formalize the definition of a LSM dataset. LSM datasets are simulated with a Monte-Carlo implementation of a dynamic pricing model. Dynamic models are parametric assumptions of the diffusion of a \emph{state vector} $S_t$, of the form:
        
        $$
            dS_t = \mu\left(S_t, t\right) dt + \sigma\left(S_t, t\right) dW_t
        $$
        
        \noindent where $S_t$ is a vector of dimension $n_0$, $\mu\left(s, t\right)$ is a vector valued function of dimension $n_0$, $\sigma\left(s, t\right)$ is a matrix valued function of dimension $n_0 \times p$ and $W_t$ is a $p$ dimensional standard Brownian motion under the \emph{pricing measure}. The number $n_0$ is called the \emph{Markov dimension} of the model, the number $p$ is called the \emph{number of factors}. Some models are non-diffusive, for example, jump diffusion models \emph{a la} Merton or rough volatility models \emph{ a la} Gatheral. All the arguments of this note carry over to more general models, but in the interest of concision and simplicity, we only consider diffusions in the exposition. Dynamic models are implemented in Monte-Carlo simulations, e.g. with Euler's scheme:
        
        $$
            S_{T_{j+1}}^{\left( i \right)} = S_{T_{j}}^{\left( i \right)} + \mu\left(S_{T_{j}}^{\left( i \right)}, T_{j}\right) \left( T_{j+1} - T_{j}\right) + \sigma\left(S_{T_{j}}^{\left( i \right)}, T_{j}\right) \sqrt{T_{j+1} - T_{j}} N_j^{\left( i \right)}
        $$
        
        \noindent where $i$ is the index of the path, $j$ is the index of the time step and the $N_j\left( i \right)$ are independent Gaussian vectors in dimension $p$.
        
        The definition of the state vector $S_t$ depends on the model. In Black and Scholes or local volatility extensions \emph{a la} Dupire, the state is the underlying asset price. With stochastic volatility models like SABR or Heston, the bi-dimensional state $S_t = \left(s_t, \sigma_t \right)$ is the pair (current asset price, current volatility). In Hull and White / Cheyette interest rate models, the state is a low dimensional latent representation of the yield curve. In general Heath-Jarrow-Morton / Libor Market models, the state is the collection of all forward rates in the yield curve. 
        
        We call \emph{model state} on date $t$ the state vector $S_{t}$ of the model on this date. 
        
        \subsubsection{Transaction state}
        
        Derivatives transactions also carry a state, in the sense that the transactions evolve and mutate during their lifetime. The state of a barrier option depends on whether the barrier was hit in the past. The state of a Bermudan swaption depends on whether it was exercised. Even the state of a swap depends on the coupons fixed in the past and not yet paid. European options don't carry state until expiry, but then, they may exercise into an underlying schedule of cashflows. 
        
        We denote $U_t$ the transaction state at time $t$ and $n_1$ its dimension. For a barrier option, the transaction state is of dimension one and contains the indicator of having hit the barrier prior to $t$. For a real-world trading book, the dimension $n_1$ may be in the thousands and it may be necessary to split the book to avoid dimension overload. The transaction state is simulated together with the model state in a Monte-Carlo implementation. In a system where event driven cashflows are scripted, the transaction state $U_t$ is also the script state, i.e. the collection of variables in the script evaluated over the Monte-Carlo path up to time $t$.
        
        \subsubsection{Training inputs}
        
        The exercise is to learn the pricing function for a given transaction or a set of transactions, in a given model, on a given date $T_1 \geq 0$, sometimes called  the \emph{exposure date} or \emph{horizon date}. The price evidently depends on both the state of the model $S_{T_1}$ and the state of the transaction $U_{T_1}$. The concatenation of these two vectors $X_{T_1} = \left [ S_{T_1} , U_{T_1} \right]$ constitute the \emph{complete Markov state} of the system, in the sense that the true price of transactions at $T_1$ are deterministic (but unknown) functions of $X_{T_1}$.
        
        The dimension of the state vector is $n_0 + n_1 = n$.
        
        The training inputs are a collection of examples $X^{\left( i \right)}$ of the Markov state $X_{T_1}$ in dimension $n$. They may be sampled by Monte-Carlo simulation between today ($T_0 = 0$) and $T_1$, or otherwise. The distribution of $X$ in the training set should reflect the intended use of the trained approximation. For example, in the context of value at risk (VAR) or expected loss (FRTB), we need an accurate approximation in extreme scenarios, hence, we need them well represented in the training set, e.g. with a Monte-Carlo simulation with increased volatility. In low dimension, the training states $X^{\left( i \right)}$ may be put on a regular grid over a relevant domain. In higher dimension, they may be sampled over a relevant domain with a low discrepancy sequence like Sobol. When the exposure date $T_1$ is today or close, sampling $X_{T_1}$ with Monte-Carlo is nonsensical, an appropriate sampling distribution must be applied depending on context. 
        
    \subsection{Pricing}
        
        \subsubsection{Cashflows and transactions}
        
        A cashflow $CF_k$ paid at time $T_k$ is formally defined as a $T_k$ measurable random variable. This means that the cashflow is revealed on or before its payment date. In the world described by the model, this is a \emph{functional} of the path of the state vector $S_t$ from $T_0 = 0$ to the payment date $T_k$ and may be simulated by Monte-Carlo.
        
        A transaction is a collection of cashflows $CF_1, ..., CF_K$. A European call of strike $K$ expiring at $T$ is a unique cashflow, paid at $T$, defined as $\left(s_T - K \right)^+$. A barrier option also defines a unique cashflow: 
        
        $$
            1_{\left\{ max \left( s_u, 0 \leq u \leq T \right) < B \right\}}\left(s_T - K \right)^+
        $$ 
        
        An interest rate swap defines a schedule of cashflows paid on its fixed leg 
        and another one paid on its floating leg. Scripting conveniently and consistently describes all cashflows, as functional of market variables, in a language purposely designed for this purpose. 
        
        A netting set or trading book is a collection of transactions, hence, ultimately, a collection of cashflows. In what follows, the word 'transaction' refers to arbitrary collection of cashflows, maybe netting sets or trading books. The payment date of the last cashflow is called the \emph{maturity} of the transaction and denoted $T_2$.
        
        \subsubsection{Payoffs}
        
        The payoff of a transaction is defined as the \emph{discounted sum of all its cashflows}:
        
        $$
            \pi = \sum_{k=1}^{K} CF_k
        $$
        
        \noindent hence, the payoff is a $T_2$ measurable random variable, which can be sampled by Monte-Carlo simulation.
        
        For the purpose of learning the pricing function of a transaction on an exposure date $T_1$, we only consider cashflows \emph{after} $T_1$, and discount them to the exposure date. In the interest of simplicity, we incorporate discounting to $T_1$ in the functional definition of the cashflows. Hence:
        
        $$
            \pi = \sum_{T_k > T_1} CF_k
        $$
        
        The payoff is still a $T_2$ measurable random variable. It can be sampled by Monte-Carlo simulation \emph{conditional to state $X_{T_1} = \left [ S_{T_1} , U_{T_1} \right]$ at $T_1$} by seeding the simulation with state $X_{T_1}$ at $T_1$ and simulating up to $T_2$. 
        
        \subsubsection{Pricing}
        
        Assuming a complete, arbitrage-free model, we immediately get the price of the transaction from the fundamental theorem of asset pricing:
        
        $$
            V_{T_1} = E \left[ \pi | F_{T_1} \right]
        $$
        
        \noindent where expectations are taken in the pricing measure defined by the model and $ F_{T_1}$ is the filtration at $T_1$ (loosely speaking, the information available at $T_1$). Since by assumption $X_{T_1} = \left [ S_{T_1} , U_{T_1} \right]$ is the complete Markov state of the system at $T_1$:
        
        $$
            V_{T_1} = E \left[ \pi | X_{T_1} \right] = h \left( X_{T_1} \right)
        $$
        
        Hence, the true price is a deterministic (but unknown) function $h$ of the Markov state.

        \subsubsection{Training labels}
        
        We see that the price corresponding to the input example $X^{\left( i \right)}$ is:
        
        $$
            V^{\left( i \right)} = E \left[ \pi | X_{T_1} = X^{\left( i \right)} \right] 
        $$
        
        \noindent and that its computation, in the general case, involves averaging payoffs over a number of Monte-Carlo simulations from $T_1$ to $T_2$, all identically seeded with $X_{T_1} = X^{\left( i \right)}$. This is also called \emph{nested simulations} because a set of simulations is necessary to compute the value of each example, the initial states having themselves been sampled somehow. If the initial states were sampled with Monte-Carlo simulations, they are called \emph{outer simulations}. Hence, we have \emph{simulations within simulations}, an extremely costly and inefficient procedure\footnote{Although, we can use nested simulations as a reference to measure performance, as we did in the working paper, sections 3.2 and 3.3. In production, nested simulations may be used to regularly double check numbers. }. 
        
        Instead, for each example $i$, we draw one single payoff $\pi^{\left( i \right)}$ from its distribution conditional to $X_{T_1} = X^{\left( i \right)}$, by simulation of one Monte-Carlo path from $T_1$ to $T_2$, seeded with $X^{\left( i \right)}$ at $T_1$. The labels in our dataset correspond to these random draws:
        
        $$
            Y^{\left( i \right)} \xleftarrow[]{sample} \pi_{T_2} | \left\{X_{T_1} = X^{\left( i \right)} \right\}
        $$

        Notice (dropping the condition to $X_{T_1} = X^{\left( i \right)}$ to simplify notations) that, while labels no longer correspond to true prices, they are \emph{unbiased} (if noisy) estimates of true prices.
        
        $$
            E \left[ Y^{\left( i \right)} \right] =  E \left[ \pi^{\left( i \right)} \right] = V^{\left( i \right)}
        $$
        
        \noindent in other terms:
        
        $$
            Y^{\left( i \right)} = V^{\left( i \right)} + \epsilon^{\left( i \right)} = h \left( X^{\left( i \right)} \right) + \epsilon^{\left( i \right)}
        $$

        \noindent where the $\epsilon^{\left( i \right)}$ are independent noise with $ E \left[ \epsilon {\left( i \right)}\right] = 0$. This is why universal approximators trained on LSM datasets converge to true prices despite having never seen one.
        
\section{Machine learning with LSM datasets}

    \subsection{Universal approximators}

    Having simulated a training set of examples $X^{\left( i \right)}, Y^{\left( i \right)}$ we proceed to train approximators, defined as functions $\hat{h} \left( x, w \right)$ of the input vector $x$ of dimension $n$, parameterized by a vector $w$ of \emph{learnable weights} of dimension $d$. This is a general definition of approximators. In classic regression, $w$ are the regression weights, often denoted $\beta$. In a neural network, $w$ is the collection of all connection matrices $W^{\left[ l \right]}$ and bias vectors $b^{\left[ l \right]}$ in the multiple layers $l=1, ..., L$ of the network.
    
    The \emph{capacity} of the approximator is an informal measure of both its computational complexity and its ability to approximate functions by matching discrete sets of datapoints. A classic formal definition of capacity is the Vapnik-Chervonenkis dimension, defined as the largest number of arbitrary datapoints the approximator can match exactly. We settle for a weaker definition of capacity as the number $d$ of learnable parameters, sufficient for our purpose.
    
    A \emph{universal} approximator is one guaranteed to approximate any function to arbitrary accuracy when its capacity is grown to infinity. Examples of universal approximator include classic linear regression, as long as the regression functions form a complete basis of the function space. Polynomial, harmonic (Fourier) and radial basis regressors are all universal approximators. Famously, neural networks are universal approximators too, a result known as the Universal Approximation Theorem.
    
    \subsection{LSM approximation theorem}
    
    Training an approximator means setting the value of its learnable parameters $w$ in order to minimize a \emph{cost function}, generally the mean square error (MSE) between the approximations and labels over a training set of $m$ examples:
    
    $$
        w = {arg min}_w MSE = \frac{1}{m} \sum_{i=1}^m \left[ \hat{h} \left( X^{\left( i \right)}, w \right) - Y^{\left( i \right)} \right]^2
    $$

    The following theorem justifies the practice of training approximators on LSM datasets:    

    \begin{quote}
    
    A universal approximator $\hat{f}$ trained by minimization of the MSE over a training set $X^{\left( i \right)}, Y^{\left( i \right)}$ of independent examples of Markov states at $T_1$ coupled with conditional sample payoffs at $T_2$, converges to the true pricing function
    
    $$
        h \left( x \right) = E \left[ \pi | X_{T_1} = x \right] 
    $$
    
    \noindent when the size $m$ of the training set and the capacity $d$ of the approximator both grow to infinity.
            
    \end{quote}
    
    We provide a sketch of proof, skipping important mathematical technicalities to highlight intuitions and important properties.
    
    First, notice that the training set consists in $m$ independent, identically distributed realizations of the couple $X, Y$ where $X$ is the Markov state at $T_1$, sampled from a distribution reflecting the intended application of the approximator, and $Y|X$ is the conditional payoff at $T_2$, sampled from the pricing distribution defined by the model and sampled by conditional Monte-Carlo simulation.
    
    Hence, the true pricing function $h$ satisfies:
    
    $$
        h \left( X \right) = E \left[ Y | X  \right] 
    $$

    By definition, the conditional expectation $ E \left[ Y | X  \right] $ is the function of $X$ closest to $Y$ in $L^2$:
    
    $$
        E \left[ Y | X  \right] \equiv min_{g \in L^2 \left(X\right)} ||g\left(X \right) - Y||_2^2
    $$
    
    Hence, pricing can be framed as an optimization problem in the space of functions. By universal approximation property:
    
    $$
        min_{w} ||\hat{h}\left(X, w \right) - Y||_2^2 \to min_{g \in L^2 \left(X\right)} ||g\left(X \right) - Y||_2^2
    $$
    
    \noindent when the capacity $d$ grows to infinity, and:
    
    $$
        MSE \to ||\hat{h}\left(X, w \right) - Y||_2^2 
    $$
    
    \noindent when $m$ grows to infinity, by assumption of an IID training set, sampled from the correct distributions. Hence:
    
    $$
        \hat{h} \left(X, min_{w} MSE\right) \to h \left( X \right)
    $$
    
    \noindent when both $m$ and $d$ grow to infinity. This is the theoretical basis for training machine learning models on LSM samples, and it applies to all universal approximators, including neural networks. This is why regression or neural networks trained on samples 'magically' converge to the correct pricing function, as observed e.g. in our demonstration notebook with European calls in Black and Scholes and basket options in Bachelier. The theorem is general and equally guarantees convergence for arbitrary (complete and arbitrage-free) models and schedules of cashflows. 
    
\section{Differential Machine Learning with LSM datasets}    
    
    \subsection{Pathwise differentials}
    
    By definition, pathwise differentials $\partial \pi / \partial X_{T_1} $ are $T_2$ measurable random variables equal to the gradient of the payoff at $T_2$ wrt the state variables at $T_1$. 
    
    For example, for a European call in Black and Scholes, pathwise derivatives are equal to:
    
    $$
        \frac{\partial \pi}{\partial S_{T_1}} = \frac{\partial \left( s_{T_2} - K\right)^+}{\partial s_{T_1}} = \frac{\partial \left( s_{T_2} - K\right)^+}{\partial s_{T_2}} \frac{\partial s_{T_2}}{\partial s_{T_1}} = 1_{\left\{ s_{T_2} > K \right\}} \frac{s_{T_2}}{s_{T_1}}
    $$
    
    In a general context, pathwise differentials are conveniently and efficiently computed with automatic adjoint differentiation (AAD) over Monte-Carlo paths as explained in the founding paper \emph{Smoking Adjoints} (Giles and Glasserman, Risk 2006) and the vast amount of literature that followed. We posted a video tutorial explaining the main ideas in 15 minutes\footnote{ \url{https://www.youtube.com/watch?v=IcQkwgPwfm4}}.
    
    Pathwise differentials are not well defined for discontinuous cashflows, like digitals or barriers. This is classically resolved by \emph{smoothing}, i.e. the replacement of discontinuous cashflows with close continuous approximations. Digitals are typically represented as tight call spreads, and barriers are represented as \emph{soft barriers}. Smoothing has been a standard practice on Derivatives trading desks for several decades. For an overview of smoothing, including generalization in terms of \emph{fuzzy logic} and a systematic smoothing algorithm, see our presentation\footnote{\url{https://www.slideshare.net/AntoineSavine/stabilise-risks-of-discontinuous-payoffs-with-fuzzy-logic}}. 
    
    Provided that all cashflows are differentiable by smoothing (and some additional, generally satisfied technical requirements), the expectation and differentiation operators commute so that true risks are (conditional) expectations of pathwise differentials:

    $$
        \text {if } h \left( x \right) = E \left[ \pi | X_{T_1} = x \right] \text { then } \frac {\partial h \left( x \right)} {\partial x} = E \left[ \frac {\partial \pi} {\partial X_{T_1}} | X_{T_1} = x \right] 
    $$
    
    This theorem is demonstrated in stochastic literature, the most general demonstration being found in \emph{Functional Ito Calculus}, also called \emph{Dupire Calculus}, see Quantitative Finance Volume 19, 2019, Issue 5. It also applies to pathwise differentials \emph{wrt model parameters}, and justifies the practice of Monte-Carlo risk reports by averaging pathwise derivatives.  
    
    \subsection{Training on pathwise differentials}
    
    LSM datasets consist of inputs $X^{\left( i \right)} = X_{T_1}^{\left( i \right)}$ with labels $Y^{\left( i \right)} = \pi^{\left( i \right)}$. Pathwise differentials are therefore the gradients of labels $Y^{\left( i \right)}$ wrt inputs $X^{\left( i \right)}$. The main proposition of the working paper is to augment training datasets with those differentials and implement an adequate training on the augmented dataset, with the result of vastly improved approximation performance.
    
    Suppose first that we are training an approximator on differentials alone:
    
    $$
        w = {arg min}_w MSE = \frac{1}{m} \sum_{i=1}^m || \frac {\partial \hat{h} \left( X^{\left( i \right)}, w \right)}{\partial X^{\left( i \right)}} - \frac{\partial Y^{\left( i \right)}} {\partial X^{\left( i \right)}} ||^2
    $$    
    
    \noindent with predicted derivatives on the left hand side (LHS) and differential labels on the right hand side (RHS). Note that the LHS \emph{is} the predicted sensitivity $\partial \hat h\left(X^{\left(i\right)}\right) / \partial X^{\left(i\right)}  $ but the RHS is \emph{not} the true sensitivity $\partial h\left(X^{\left(i\right)}\right) / \partial X^{\left(i\right)}  $. It is the pathwise differential, a random variable with expectation the true sensitivity and additional sampling noise. 
    
    We have already seen this exact same situation while training approximators on LSM samples, and demonstrated that the trained approximator converges to the true conditional expectation, in this case, the expectation of pathwise differentials, a.k.a. the true risk sensitivities. 
    
    The trained approximator will therefore converge to a function $\hat{h}$ with all the same differentials as the true pricing function $h$. It follows that on convergence $\hat{h} = h$ modulo an additive constant $c$, trivially computed at the term of training by matching means: 
    
    $$
        c = \frac {1} {m} \sum_{i=1}^m \left[ Y^{\left( i \right)} - \hat{h} \left( X^{\left( i \right)} \right) \right]
    $$

\section*{Conclusion}
    
We reviewed the details of LSM simulation framed in machine learning terms, and demonstrated that training approximators on LSM datasets effectively converges to the true pricing functions. We then proceeded to demonstrate that the same is true of differential training, i.e. training approximators on pathwise differentials also converges to the true pricing functions. 

These are asymptotic results. They justify standard practices and guarantee consistence and meaningfulness of classical and differential training on LSM datasets, classical or augmented e.g. with AAD. They don't say anything about speed of convergence. In practicular, they don't provide a quantification of errors with finite capacity $d$ and finite datasets of size $m$. They don't explain the vastly improved performance of differential training, consistently observed across examples of practical relevance in the working paper. Both methods have the same asymptotic guarantees, where they differ is in the magnitude of errors with finite capacity and size. To quantify those is a more complex problem and a topic for further research.
\pagebreak 

\chapter{Taking the First Step : Differential PCA}
\counterwithout{section}{chapter}
\setcounter{section}{0}
\label{app2}
\section*{Introduction}

We review traditional data preparation in deep learning (DL) including principal component analysis (PCA), which effectively performs orthonormal transformation of input data, filtering constant and redundant inputs, and enabling more effective training of neural networks (NN). Of course, PCA is also useful in its own right, providing a lower dimensional latent representation of data variation along orthogonal axes. 

In the context of \emph{differential} DL, training data also contains differential labels (differentials of training labels wrt training inputs, computed e.g. with automatic adjoint differentiation -AAD- as explained in the working paper), and thus requires additional preprocessing. 

We will see that differential labels also enable remarkably effective data preparation, which we call \emph{differential PCA}. Like classic PCA, differential PCA provides a hierarchical, orthogonal representation of data. Unlike classic PCA, differential PCA represents input data in terms how it affects the target measured by training labels, a notion we call \emph{relevance}. For this reason, differential PCA may be safely applied to aggressively remove irrelevant factors and considerably reduce dimension.

In the context of data generated by financial Monte-Carlo paths, differential PCA exhibits the principal risk factors of the target transaction or trading book from data alone. It is therefore a very useful algorithm on its own right, besides its effectiveness preparing data for training NN.

The first section describes and justifies elementary data preparation, as implemented in the demonstration notebook \emph{DifferentialML.ipynb} on  \url{https://github.com/differential-machine-learning}. Section 2 discusses the mechanism, benefits and limits of classic PCA. Section 3 introduces and derives differential PCA and discusses the details of its implementation and benefits. Section 4 brings it all together in pseudocode.

\section{Elementary data preparation}

Dataset normalization is known as a crucial, if somewhat mundane preparation step in deep learning (DL), highlighted in all DL textbooks and manuals. Recall from the working paper that we are working with augmented datasets:

$$
    X ^{\left( i \right)} : \text {inputs, } Y ^{\left( i \right)} : \text {labels, and } Z ^{\left( i \right)} = \frac{\partial Y ^{\left( i \right)}}{\partial X ^{\left( i \right)}} : \text {differential labels }
$$

\noindent with $m$ labels in dimension 1 and $m$ inputs and $m$ differentials in dimension $n$, stacked in $m$ rows in the matrices $X$, $Y$ and $Z$. In the context of financial Monte-Carlo simulations \emph{a la} Longstaff-Schwartz, inputs are Markov states on a \emph{horizon date} $T_1 \geq 0$, labels are payoffs sampled on a later date $T_2$ and differentials are pathwise derivatives, produced with AAD. 

The normalization of augmented datasets must take additional steps compared to conventional preparation of classic datasets consisting of only inputs and labels.

\subsection{Taking the first (and last) step}

A first, trivial observation is that the scale of labels $Y^{\left( i \right)}$ carries over to the gradients of the cost functions and the size of gradient descent optimization steps. To avoid manual scaling of learning rate, gradient descent and variants are best implemented with labels normalized by subtraction of mean and division by standard deviation. This is the case for all models trained with gradient descent, including classic regression in high dimension where the closed form solution is intractable.

Contrarily to classic regression, training a neural network is a nonconvex problem, hence, its result is sensitive to the starting point. Correctly seeding connection weights is therefore a crucial step for successful training. The best practice \emph{Xavier-Glorot} initialization provides a powerful seeding heuristic, implemented in modern frameworks like TensorFlow. It is based on the implicit assumption that the units in the network, including inputs, are centred and orthonormal. It therefore performs best when the inputs are at the very least normalized by mean and standard deviation, and ideally orthogonal. This is specific to neural networks. Training classic regression models, analytically or numerically, is a convex problem, so there is no need to normalize inputs or seed weights in a particular manner.

Training deep learning models therefore always starts with a normalization step and ends with a 'un-normalization step' where predictions are scaled back to original units. Those first and last step may be seen, and implemented, as additional layers in the network with fixed (non learnable) weights. They may even be merged in the input and output layer of the network for maximum efficiency. In this document, we present normalization as a preprocessing step in the interest of simplicity.

\subsubsection{First step}

We implemented basic preprocessing in the demonstration notebook:

$$
    \tilde Y ^{\left( i \right)} = \frac {Y ^{\left( i \right)} - \mu_Y} {\sigma_Y} \text{ and } \tilde X_j ^{\left( i \right)} = \frac {X_j ^{\left( i \right)} - \mu_j} {\sigma_j}
$$

\noindent where 

$$
    \mu_Y = \frac{1}{m} \sum_{i=1}^{m} Y ^{\left( i \right)}  \text{ and } \mu_j = \frac{1}{m} \sum_{i=1}^{m} X_j ^{\left( i \right)}
$$

\noindent and similarly for standard deviations of labels ${\sigma_Y}$ and inputs ${\sigma_j}$. 

The differentials computed by the prediction model (e.g. the twin network of the working paper) are:

$$
    \frac {\partial \tilde Y} {\partial \tilde X_j} = \frac {\sigma_j} {\sigma_Y} \frac {\partial Y} {\partial  X_j}
$$

\noindent hence, we adjust differential labels accordingly:

$$
    \tilde Z_j ^{\left( i \right)} = \frac {\sigma_j} {\sigma_Y} Z_j ^{\left( i \right)}
$$

\subsubsection{Training step}

The value labels $\tilde Y$ are centred and unit scaled but the differentials labels $\tilde Z$ are not, they are merely re-expressed in units of 'standard deviations of $Y$ per standard deviation of $X_j$'. To avoid summing apples and oranges in the combined cost function as commented in the working paper, we scale cost as follows:

\[
    C = \frac{1}{m} \sum_{i=1}^{m} \left[ \hat Y^{\left( i \right)} \left( w\right) - \tilde Y ^{\left( i \right)} \right]^2 + \frac{1}{m} \sum_{i=1}^{m} \sum_{j=1}^{n} \frac{1}{||\tilde Z_j||^2}  \left[ \hat Z_j^{\left( i \right)} \left( w\right) - \tilde Z ^{\left( i \right)} \right]^2  \tag{C} \label{eq:cost}
\]

\noindent and proceed to find the optimal biases and connection weights by minimization of $C$ in $w$.

\subsubsection{Last step}

The trained model $\tilde f$ expects normalized inputs and predicts a normalized value, along with its gradient to the normalized inputs. Those results must be scaled back to original units:

$$
    f \left( x \right) = \mu_Y + \sigma_Y \tilde f \left( \tilde x \right) = \mu_Y + \sigma_Y \tilde f \left( \frac {x - \mu_X} {\sigma_X} \right)
$$

\noindent where we divided two row vectors to mean elementwise division, and:

$$
    \frac {\partial f \left( x \right)} {\partial x_j} = \frac {\sigma_Y} {\sigma_j}  \frac {\partial \tilde f \left( \tilde x \right)} {\partial \tilde x_j}
$$

\subsection{Limitations}

Basic data normalization is sufficient for textbook examples but more thorough processing is necessary in production, where datasets generated by arbitrary schedules of cashflows simulated in arbitrary models may contain a mass of constant, redundant of irrelevant inputs. Although neural networks are supposed to correctly sort data  and identify relevant features during training\footnote{SVD regression performs a similar task in the context of classic regression, see note on differential regression.}, in practice, nonconvex optimization is much more reliable when at least \emph{linear} redundancies and irrelevances are filtered in a preprocessing step, lifting those concerns from the training algorithm and letting it focus on the extraction of \emph{nonlinear} features. 

In addition, it is best, although not strictly necessary, to train on orthogonal inputs. As it is well known, normalization and orthogonalization of input data, along with filtering of constant and linearly redundant inputs, is all jointly performed in a principled manner by eigenvalue decomposition of the input covariance matrix, in a classic procedure called principle component analysis or PCA. 

\section{Principal Component Analysis}

\subsection{Mechanism}

We briefly recall the mechanism of data preparation with classic PCA. First, normalize labels and center inputs:

$$
    \tilde Y ^{\left( i \right)} = \frac {Y ^{\left( i \right)} - \mu_Y} {\sigma_Y} \text{ and } X_j ^{\left( i \right)} \equiv X_j ^{\left( i \right)} - \mu_j
$$

\noindent i.e. what we now call $X$ is the matrix of \emph{centred} inputs. Perform its eigenvalue decomposition:

$$
    \frac{1}{m} X^T X = P D P^T
$$

\noindent where $P$ is the orthonormal $n \times n$ matrix of eigenvectors (in columns) and $D$ is the diagonal matrix of eigenvalues $D_{jj}$. 

Filter numerically constant or redundant inputs identified by eigenvalues $D_{jj}$ lower than a threshold $\epsilon$. The filter matrix $F$ has $n$ rows and $\tilde n \leq n$ columns and is obtained from the identity matrix $I_n$ by removal of columns corresponding to insignificant eigenvalues $D_{jj}$. Denote:

$$
    \tilde D  = F^T D F \text { and } \tilde P = P F
$$

\noindent the reduced eigenvalue and eigenvector matrices of respective shapes $\tilde n \times \tilde n$ and $n \times \tilde n$, and apply the following linear transformation to centred input data:

$$
    \tilde X = X \tilde P \tilde D^{-\frac{1}{2}}
$$

The transformed data has shape $m \times \tilde n$, with constant and linearly redundant columns filtered out. It is evidently centred, and easily proved orthonormal:

    \begin{eqnarray*}
    \frac 1 m \tilde X^T \tilde  X & = &  \tilde  D^{-\frac 12} \tilde P^T \left( \frac 1 m  X^T  X \right) \tilde P \tilde D^{-\frac 12} \\
    & = & \left( F^T  D F \right)^{-\frac 12} \left(P F\right)^T \left( \frac 1 m X^T X \right) \left(P F \right)\left(F^T D F\right)^{-\frac 12} \\
    & = & \left( F^T D F \right)^{-\frac 12} F^T P^T P D P^T P F \left( F^T D F \right)^{-\frac 12}\\
     & = & \left( F^T D F \right)^{-\frac 12}\left( F^T D F \right) \left( F^T D F \right)^{ -\frac 12}\\
     & = & I_{\tilde n}
     \end{eqnarray*}
     
Note for what follows that orthonormal property is preserved by rotation, i.e. right product by any orthonormal matrix $Q$:

    $$
        \frac 1m \left(\tilde X Q \right)^T \left( \tilde X Q \right) = Q^T \left( \frac 1m {\tilde X^T \tilde X} \right) Q = Q^T I_{\tilde n} Q = Q^T Q = I_{\tilde n}
    $$
    
To update differential labels, we apply a result from elementary multivariate calculus:

\begin{quote}
\label{lemma}
    Given two row vectors $A$ and $B$ in dimension $p$ and a square non singular matrix $M$ of shape $p \times p$ such that $B = A M$, and $y=g \left( A \right) = h \left( B \right) $ a scalar, then:
    
    $$ 
        \frac {\partial y} {\partial B} = \frac {\partial y} {\partial A}  M^{-T}
    $$
    The proof is left as an exercise.
\end{quote}

It follows that:

    \begin{eqnarray*}
    \tilde Z ^{\left( i \right)}    & = & \frac {\partial \tilde Y^{\left( i \right)}} {\partial \tilde X^{\left( i \right)}} \\
                                    & = & \frac {\partial \left( Y^{\left( i \right)} / \sigma_Y \right)} {\partial \left( X^{\left( i \right)}  \tilde P \tilde D^{-\frac{1}{2}} \right)} \\
                                    & = & \frac 1 {\sigma_Y}  \frac {\partial Y^{\left( i \right)} } {\partial X^{\left( i \right)} }  \left( \tilde P \tilde D^{-\frac{1}{2}}  \right)^{-T}          \\
                                    & = & \frac 1 {\sigma_Y} Z ^{\left( i \right)}      \tilde P \tilde D^{\frac{1}{2}}  \\
    \end{eqnarray*}
    
Or the other way around:

$$
    \frac {\partial y} {\partial x} = \sigma_Y \frac {\partial \tilde y} {\partial \tilde x} \tilde D^{- \frac 1 2 } \tilde P^T 
$$

We therefore train the ML model $\tilde f$ more effectively on transformed data:

$$
    \tilde X = \left( X - \mu_X \right) \tilde P \tilde D^{-\frac{1}{2}}
    \text{ , } \tilde Y  = \frac {Y  - \mu_Y} {\sigma_Y} 
    \text{ and } \tilde Z = \frac 1 {\sigma_Y} Z \tilde P \tilde D^{\frac{1}{2}} 
$$

\noindent by minimization of the the cost function \eqref{eq:cost} in the learnable weights. The trained model $\tilde f$ takes inputs $\tilde x$ in the tilde basis and predicts normalized values $\tilde y$ and differentials ${\partial \tilde y} / {\partial \tilde x}$. Finally, we translate predictions back in the original units:

$$
    f \left( x \right) = \mu_Y + \sigma_Y \tilde f \left( \tilde x  \right) \text { and } 
    \frac {\partial f \left( x \right)} {\partial x} = \sigma_Y \frac {\partial \tilde f \left( \tilde x  \right)} {\partial \left( \tilde x  \right)} \tilde D^{- \frac 1 2 } \tilde P^T \text { where } \tilde x = (x - \mu_x) \tilde P \tilde D^{-\frac 1 2}
$$

PCA performs an orthonormal transformation of input data, removing constant and linearly redundant columns, effectively cleaning data to facilitate training of NN. PCA is also useful in its own right. It identifies the main axes of variation of a data matrix and may result in a lower dimensional latent representation, with many applications in finance and elsewhere, covered in vast amounts of classic literature.

PCA is limited to \emph{linear} transformation and filtering of \emph{linearly} redundant inputs. A nonlinear extension is given by autoencoders (AE), a special breed of neural networks with bottleneck layers. AE are to PCA what neural networks are to regression, a powerful extension able to identify lower dimensional \emph{nonlinear} latent representations, at the cost of nonconvex numerical optimization. Therefore, AE themselves require careful data preparation and are not well suited to prepare data for training other DL models.

\subsection{Limitations}

\subsubsection{Further processing required}

In the context of a differential dataset, we cannot stop preprocessing with PCA. Recall, we train by minimization of the cost function \eqref{eq:cost}, where derivative errors are scaled by the size of differential labels. We will experience numerical instabilities when some differential columns are identically zero or numerically insignificant. This means the corresponding inputs are \emph{irrelevant} in the sense that they don't affect labels in any of the training examples. They really should not be part the training set, all they do is unnecessarily increase dimension, confuse optimizers and cause numerical errors. But PCA cannot eliminate them because it operates on inputs alone and disregards labels and how inputs affect them. \emph{PCA ignores relevance}.

Irrelevances may even appear in the orthogonal basis, even when inputs looked all relevant in the original basis. To see that clearly, consider a simple example in dimension 2, where $X_1$ and $X_2$ are sampled from 2 standard Gaussian distributions with correlation $1/2$ and $Y = X_2 - X_1 + \text{noise}$. Differential labels are constant across examples with $Z_1 = -1$ and $Z_2 = 1$. Both differentials are clearly nonzero and both inputs appear to be relevant. PCA projects data on orthonormal axes $\tilde X_1 = \left( X_1 + X_2 \right) / \sqrt{2}$ and $\tilde X_2 = \left( - X_1 + X_2\right) / \sqrt{2}$ with eigenvalues $3/2$ and $1/2$, and:

$$
    \tilde Z_1 = \frac {\partial Y} {\partial \tilde X_1} = \sqrt{2} \frac {\partial \left( X_2 - X_1\right)} {\partial \left( X_2 + X_1 \right)} = 0
$$

\noindent so after PCA transformation, one of the columns clearly appears irrelevant. Note that this is a coincidence, we would not see that if correlation between $X_1$ and $X_2$ were different from $1/2$. PCA is not able to identify axes of relevance, it only identifies axes of variation. By doing so, it \emph{may} accidentally land on axes with zero or insignificant relevance.

It appears from this example that, not only further processing is necessary, but also, desirable to eliminate irrelevant inputs and combinations of inputs in the same way that PCA eliminated constant and redundant inputs. Note that we don't want to \emph{replace} PCA. We want to train on orthonormal inputs and filter constants and redundancies. What we want is \emph{combine} PCA with a similar treatment of relevance.

\subsubsection{Limited dimension reduction}

The eventual amount of dimension reduction PCA can provide is limited, precisely because it ignores relevance. Consider the problem of a basket option in a correlated Bachelier model, as in the section 3.1 of the working paper. The states $X$ are realizations of the $n$ stock prices at $T_1$ and the labels $Y$ are option payoffs, conditionally sampled at $T_2$. Recall that the price at $T_1$ of a basket option expiring at $T_2$ is a nonlinear scalar function (given by Bachelier's formula) \emph{of a linear combination $X \cdot a$} of the stock prices $X$ at $T_1$, where $a$ is the vector of weights in the basket. The basket option, which payoff is measured by $Y$, is only affected (in a nonlinear manner) by \emph{one} linear risk factor $X \cdot a$ of $X$. Although the input space is in dimension $n$, the subspace of relevant risk factors is in dimension $1$. Yet, when the covariance matrix of $X$ is of full rank, PCA identifies $n$ axes of orthogonal variation. It only reduces dimension when the covariance matrix is singular, to eliminates trivially constant or redundant inputs, even in situations where dimension could be reduced by significantly larger amounts with relevance analysis.  

\subsubsection{Unsafe dimension reduction}

In addition, it is not desirable to attempt aggressively reducing dimension with PCA because it could eliminate relevant information. To see why, consider another, somewhat contrived example with 500 stocks, one of them (call it XXX)  uncorrelated with the rest with little volatility. An aggressive application of PCA would remove that stock from the orthogonal representation, even in the context of a trading book dominated by a large trade on XXX. PCA \emph{should not} be relied upon to reduce dimension, because it ignores relevance and hence, might accidentally remove important features. PCA must be applied \emph{conservatively}, with filtering threshold $\epsilon$ set to numerically insignificant eigenvalues, to only eliminate definitely constant or linearly redundant inputs.

\subsubsection{Principal components are not risk factors}

The Bachelier basket example makes it clear that the orthogonal axes of variation identified by PCA are \emph{not} risk factors. In general, PCA provides a meaningful orthonormal representation of the state vector $X$, but it doesn't say anything about the factors affecting the transaction or its cashflows measured by the labels $Y$. 

\section{Differential PCA}

\subsection{Introduction}

The question is then whether we can design an additional, similar procedure to effectively extract from simulated data the risk factors \emph{of a given transaction}, and \emph{safely} reduce dimension by removal of \emph{irrelevant} axes? We expect the algorithm to identify the basket weights as the only risk factor in the Bachelier example, and XXX alone as a major risk factor in the 500 stocks example. In the general case, we want to reliably extract a hierarchy of orthogonal risk factors and safely eliminate irrelevant directions.

To achieve this, we turn to differential labels $Z ^{\left( i \right)} = {\partial Y ^{\left( i \right)}} / {\partial X ^{\left( i \right)}}$, which, in the context of simulated financial data, are either risk sensitivities or \emph{pathwise differentials}\footnote{Depending on whether labels are ground truth or sampled, see working paper.}. The main proposition of differential machine learning is to leverage differential labels computed e.g. with AAD, and we have seen their effectiveness for approximation by neural networks (main article) or classic regression (appendices). Here, we will see that they also apply in the context of PCA, not to improve it, but to combine it with an additional procedure, which we call 'differential PCA', capable of exhibiting orthogonal risk factors and safely removing irrelevant combinations of inputs.  

As a data preparation step, differential PCA may significantly reduce dimension, enabling faster, more reliable training of neural networks, and a reduced sensitivity to seeding and hyperparameters. In particular, We will see that differential PCA reduces dimension while \emph{preserving} orthonormality of inputs from a prior PCA step. 

In its own right, differential PCA reliably identifies risk factors from simulated data. Like traditional PCA, it only extracts \emph{linear} factors, but unlike PCA, it analyses and transforms on data through the lens of \emph{relevance}.

\subsection{Derivation}

Start with a dataset:

$$
    X : \text {inputs } X ^{\left( i \right)} \text{ , } Y : \text {labels } Y ^{\left( i \right)} \text {, and } Z: \text{differentials } Z ^{\left( i \right)} = \frac{\partial Y ^{\left( i \right)}}{\partial X ^{\left( i \right)}} \text { stacked in rows in matrices } X, Y, Z
$$

\noindent possibly orthonormal by prior PCA, with differentials appropriately adjusted and constant and redundant inputs filtered out. 

We want to apply a \emph{rotation} by right multiplication of the input matrix $X$ by an orthonormal matrix $Q$ so as to preserve orthonormality:

$$
    \tilde X = X Q
$$

\noindent so that the directional differentials $\tilde Z_j ^{\left( i \right)} = \partial Y  ^{\left( i \right)} / \partial \tilde X_j^{\left( i \right)}$ are mutually orthogonal. Recall from the lemma page \pageref{lemma}:

$$
    \tilde Z = Z Q ^{-T} = Z Q
$$

\noindent and we want $\tilde Z$ to be orthogonal, i.e.:

$$
    \frac 1 m \tilde Z^T \tilde Z = E
$$

\noindent with $E$ a diagonal matrix whose entries $E_{jj}$ are the mean norms of the columns $\tilde Z_j$ of $\tilde Z$, in other terms, the \emph{size} of differentials, also called \emph{relevance}, in the tilde basis.

Since $\tilde Z = Z Q$,

$$
    \frac 1 m \tilde Z^T \tilde Z = \frac 1 m Q^T Z^T Z Q = E
$$

\noindent or inverting:

$$
    \frac 1 m Z^T Z = Q E Q^T 
$$

\noindent and we have the remarkable solution that $Q$ and $E$ are the eigenvectors and eigenvalues of the empirical covariance matrix of derivatives labels $\left( 1/m \right) Z^T Z$.

We can proceed to eliminate irrelevant directions by right multiplication by a filter matrix on a criterion $E_{jj} > \epsilon'$. 

\begin{quote}
    Hence, differential PCA is simply PCA on differential labels.
\end{quote}

Unlike with PCA, it is safe to filter irrelevance aggressively. Assuming that a prior PCA step was performed, differentials are expressed in 'standard deviation of labels per standard deviation of inputs'. Eigenvalues $E_{jj}$ less than $10^{-4}$ reflect a sensitivity less than $10^{-2}$, where it takes more than $100$ standard deviations in the input to produce \emph{one} standard deviation in the label. The corresponding input can safely be discarded as irrelevant. It is therefore reasonable to set the filtering threshold $\epsilon'$ on differentials to $10^{-4}$ or even higher without fear of losing relevant information, whereas the PCA threshold $\epsilon$ should be near numerical zero to avoid information loss.

Note that we performed differential PCA on the \emph{noncentral} covariance matrix of derivatives. Constant derivatives correspond to linear factors, which we must consider relevant, at least for training. In order to extract nonlinear risk factors only, we could apply the same procedure with eigenvalue decomposition of the centred covariance matrix $ \frac 1 m \left(Z - \mu_Z \right)^T \left(Z - \mu_Z \right) = Q E Q^T $ instead.

\subsection{Example}

In the context of the simple example of a basket option with weights $a$ in a correlated Bachelier model, we can perform differential PCA explicitly. The input matrix $X$ of shape $m \times n$ stacks $m$ rows of examples $ X ^{\left( i \right)}$, each one a row vector of the $n$ stock prices on a horizon date $T_1$. The label vector $Y$ collects corresponding payoffs for the basket option of strike $K$, sampled on the same path on a later date $T_2$:

$$
Y ^{\left( i \right)} =  \left(S_{T_2}  ^{\left( i \right)} \cdot a - K \right)^+
$$

For simplicity, we skip the classic PCA step. The differential labels, in this simple example, are money indicators:

$$
Z ^{\left( i \right)} = \frac {\partial Y ^{\left( i \right)}} {\partial X ^{\left( i \right)}} = 1_{\left\{ S_{T_2}^{\left( i \right)} \cdot a > K \right\}} a^T 
$$

\noindent with the usual notations. Differential labels are $0_n$ on paths finishing out of the money and $a$ on paths finishing in the money. Denote $q$ the empirical proportion of paths finishing in the money. Then:

$$
    \frac 1 m Z^T Z = q a a^T 
$$

\noindent and its eigenvalue decomposition $ 1 / m Z^T Z = Q E Q^T$  is: 

$$Q = \left[ {\left[ {\frac{a}{{\left\| a \right\|}}} \right]\left[ \begin{array}{l}
0\\
...\\
0
\end{array} \right]...\left[ \begin{array}{l}
0\\
...\\
0
\end{array} \right]} \right],E = \left[ {\begin{array}{*{20}{c}}
{q\left\| a \right\|^2}&{}&{}&{}\\
{}&0&{}&{}\\
{}&{}&{...}&{}\\
{}&{}&{}&0
\end{array}} \right]$$

Hence, differential PCA gives us a single relevant risk factor, exactly corresponding to the (normalized) weights in the basket.

\section{A complete data preparation algorithm}

We conclude with a complete data processing algorithm in pseudocode, switching to adjoint notations, i.e. we denote $\bar X$ what we previously denoted $Z$. We also use subscripts to denote processing stages.
 
\begin{enumerate}
    \setcounter{enumi}{-1}
    \item Start with raw data $\underbrace {X_0}_{m \times n_0},\underbrace {Y_0}_{m \times 1},\underbrace {\bar X_0}_{m \times n_0}$. 
    
    \item Basic processing
    
    \begin{enumerate}
        \item Center inputs (but do not normalize them quite yet) with means the row vector $\mu_x$ of dimension $n_0$, computed across training examples:
        \[
            \left( X_1 \right)_i = \left( X_0 \right)_i - \mu _x
        \]
        
        \item Compute standard deviation $\sigma_y$ of labels across examplesand normalize labels: $Y_1 = \frac{Y_0 - \mu_y}{\sigma_y}$
        
        \item Update derivatives: $            \bar X_1 = \frac{\bar X_0}{\sigma_y} $
        
        \item Reverse for translating predictions: inputs must be normalized first, consistently with training inputs. The model returns a normalized prediction $\hat y_1$ and its derivatives $\hat {\bar x}_1$. Translate predictions back into original units with the reverse transformations: 
        
        $$\hat y_0 = \mu_y + \sigma_y \hat y_1 \text{ and } \hat {\bar x}_0 + \sigma_y \hat{\bar x}_1$$
        
    \end{enumerate}
    
    \item PCA
    
    \begin{enumerate}

        \item Perform eigenvalue decomposition of $\frac{X_1^T X_1}{m} = P_2D_2P_2^{-1}$.

        \item Shrink the diagonal matrix $D_2$ to dimension $n_2 \le n_1$ by removing rows and columns corresponding to numerically zero eigenvalues. Denote $F_2$ the filter matrix of shape ($n_1$, $n_2$), obtained by removal of columns in the identity matrix $I_{n_1}$ corresponding to numerically null diagonal entries of $D_2$. The reduced diagonal matrix is $\tilde D_2 = F_2^T D_2 F_2$ of size $n_2$, and the reduced eigenvector matrix is $\tilde P_2 = P_2 F_2$ of shape ($n_1$, $n_2$). Note that the eigenvalue matrix remains diagonal and the columns of the eigenvector matrix remain orthogonal and normalized after filtering. 
        
        \item Apply the orthonormal transformation:
        
        $$
        X_2 = X_1 \tilde P_2 \tilde D_2^{-\frac 12}
        $$
        
        \item Update differentials
        
        \begin{eqnarray*}
            \bar X_2 & = & \bar X_1 \tilde P_2\tilde D_2^{\frac 12}
        \end{eqnarray*}
        
        \item Note the reverse formula for prediction of derivatives: $\bar X_1  = \bar X_2 \tilde D_2^{-\frac 12}\tilde P_2^T$
        
    \end{enumerate}
    
    \item Differential PCA
    
    \begin{enumerate}
    
        \item Perform eigenvalue decomposition of:
        
        $$
        \frac{\bar X_2^T \bar X_2}{m} = P_3 D_3 P_3^{-1}
        $$
        
        \item Shrink the columns of the eigenvector matrix $P_3$ to dimension $n_3\le n_2$ by removing columns corresponding to small eigenvalues. Denote $F_3$ the corresponding filter matrix of shape ($n_2$, $n_3$). The reduced inverse eigenvector matrix is $\tilde P_3 = P_3 F_3$ of shape ($n_2$, $n_3$).
        
        \item Apply the linear transformation: 
        
        $$
        X_3 = X_2 \tilde P_3
        $$
        
        \item Update differentials: 
        
        $$
        \bar X_3  = \bar X_2 \tilde P_3
        $$
        
        \item Note reverse formula for prediction: 
        
        $$
        \bar X_2  = \bar X_3 \tilde P_3^T
        $$
    
    \end{enumerate}

    \item Train model on the dataset $X_3$, $Y_3$, $\bar X_3$, which is both orthonormal is terms of inputs $X_3$ and orthogonal in terms of directional differentials $\bar X_2$, with constant, redundant and irrelevant inputs and combinations filtered out. 
    
    \item Predict values and derivatives with the trained model $\hat y = \hat f \left( x \right)$ from raw inputs $x = x_0$:
    
    \begin{enumerate}
        
        \item Transform inputs:
        
        \begin{enumerate}
            
            \item       $ x_1  = x_0  - \mu _x $
            \item       $ x_2 = x_1 \tilde P_2 \tilde D_2^{-\frac 12} $
            \item       $ x_3 = x_2 \tilde P_3 $
            
        \end{enumerate}
    
        \item predict values:
         \begin{enumerate}
            \item       $ \hat y_3 = \hat f \left( x_3 \right)$
            \item       $ \hat y_0 = \mu_Y + \sigma_Y \hat y_3$
        \end{enumerate}
        
        \item predict derivatives:
        
         \begin{enumerate}

        \item $ \hat{ \bar x}_3 =  \frac {\partial \hat f \left( x_3 \right)} {\partial x_3}$
        \item $ \hat{ \bar x}_2 = \hat{ \bar x}_3  \tilde  P_3^T $
        \item $ \hat{ \bar x}_1 = \hat{ \bar x}_2  \tilde  \tilde D_2^{-\frac 12}\tilde P_2^T$ 
        \item $ \hat{ \bar x}_0 = \sigma_y \hat{\bar x}_1$

        \end{enumerate}            
    
    \end{enumerate}
    
\end{enumerate}

\section*{Conclusion}

We derived a complete data preparation algorithm for differential deep learning, including a standard PCA step and a differential PCA step. Standard PCA performs an orthonormal transformation of inputs and eliminates constant and redundant ones, facilitating subsequent training of neural networks. Differential PCA further rotates data to an orthogonal relevance representation and may considerably reduce dimension, in a completely safe manner, by elimination of irrelevant directions. 

Like standard PCA, differential PCA is a useful algorithm on its own right, providing a low dimensional latent representation of data on orthogonal axes of relevance. In the context of financial simulations, it computes an orthogonal hierarchy of risk factors for a given transaction. For example, we proved that differential PCA identifies basket weights as the only relevant risk factor for a basket option, from simulated data alone. 

We achieved this by leveraging information contained in the differential labels, which, in the context of simulated financial data, are  \emph{pathwise differentials} and contain a wealth of useful information. Recall that traditional risk reports are averages of pathwise differentials. Averaging, however, collapses information. For example, the risk report of a delta-hedged European call (obviously) returns zero delta, although the underlying stock price most definitely remains a relevant risk factor, affecting the trading book in a nonlinear manner, an information embedded in pathwise differentials but eliminated by averaging. Pathwise differentials are sensitivities of payoffs to state in a multitude of scenarios. They have a broader story to tell than aggregated risk reports. 

The main proposition of the article is to leverage differentials in all sort of machine learning tasks, and we have seen their effectiveness for approximation by neural networks (main article) or classic regression (appendices). Here, we have seen that they also apply in the context of PCA, not to improve it, but to combine it with an additional procedure, which we call 'differential PCA', capable of exhibiting risk factors and safely removing irrelevant combinations of inputs. As a preprocessing step, differential PCA makes a major difference for training function approximations, reducing dimension, stabilizing nonconvex numerical optimization and reducing sensitivity to initial seed and hyperparameters like neural architecture or learning rate schedule.
\pagebreak 

\chapter{Differential Regression}
\counterwithout{section}{chapter}
\setcounter{section}{0}
\label{app3}
\section*{Introduction}

Differential machine learning is presented in the paper in the context of deep learning, where its \emph{unreasonable effectiveness} is illustrated with examples picked in both real-world applications and textbooks, like the basket option in a correlated Bachelier model. 

In this appendix, we apply the same ideas in the context of classic regression on a fixed set of basis functions, and demonstrate equally remarkable results, illustrated with the same Bachelier basket example, with pricing and risk functions approximated by polynomial regression. Recall that the example from the paper is reproduced on a public notebook \url{https://github.com/differential-machine-learning/notebooks/blob/master/DifferentialML.ipynb}. We posted another notebook \emph{DifferentialRegression.ipynb} with the regression example, where the formulas of this document for standard, ridge and differential regression are implemented and compared. 

Like standard and ridge regression, differential regression is performed in closed form and lends itself to SVD stabilization. Unlike ridge regression, differential regression provides strong regularization \emph{without bias}. It follows that there is no bias-variance tradeoff with differential regression, in particular, the sensitivity to regularization strength is virtually null. As illustrated in the notebook, differential regression vastly outperforms Tikhonov regularization, even when the Tikhonov parameter is optimized by cross validation at the cost of additional data consumption. Differential regression doesn't consume additional data besides a training set augmented with differentials as explained in the paper. It doesn't necessitate additional regularization or hyperparameter optimization by cross validation.

The exercise is to perform a classic least square linear regression $\widehat Y = {\mu _Y} + \left( {\phi  - {\mu _\phi }} \right)\beta $, where the columns of $\phi  = \phi \left( X \right)$  are basis functions (e.g. monomials, excluding constant) of known inputs $X$  (also excluding constant, with examples in rows and inputs in columns), given a column vector $Y$  of the corresponding targets, where $\mu_Y$  is the mean of $Y$  and the row vector ${\mu _\phi }$  contains the means of the columns of  $\phi $. To simplify notations, we denote $\phi  \equiv \phi  - {\mu _\phi }$  and  $Y \equiv Y - {\mu _Y}$. Classic least squares finds  $\beta$ by minimization of the least square errors:
  
  $$\beta  = \arg {\min _\beta }{\left\| {Y - \phi \beta } \right\|^2}$$
  
The analytic solution, also called ‘normal equation’:
  
  $$\beta  = {\left( {{\phi ^T}\phi } \right)^{ - 1}}{\phi ^T}Y$$
  
is known to bear unstable results, the matrix  ${\phi ^T}\phi $  often being near singular (certainly so with monomials of high degree of correlated inputs). This is usually resolved with SVD regression. We prefer the (very similar) eigenvalue regression, which we recall first, and then, extend to ridge (Tikhonov) regularization and finally differential regression.
Parts 1 and 2 are summaries of classic results. Part 3 is new.
After $\beta$  is learned, the value approximation for an input row vector $x$  is given by $\widehat y = \phi \left( x \right) \beta $  and the derivative approximations are given by:

\[{\widehat y_j} = {\phi _j}\left( x \right)\beta \]
 
where subscripts denote partial derivatives to input number $j$.

\section{SVD regression, eigenvalue variant}

Perform the eigenvalue decomposition  ${\phi ^T}\phi  = PD{P^T}$. 

Denote ${D^{ - \frac{1}{2}}}$  the diagonal matrix whole diagonal elements are the elements of the diagonal matrix $D$, raised to power $-0.5$ , when they exceed a threshold (say, $10^{-8}$  times the mean trace of $D$), and zero otherwise.

Denote $\widetilde \phi  = \phi P{D^{ - \frac{1}{2}}}$  and perform the least square minimization in the orthonormal basis:
 
 $$\widetilde \beta  = \arg {\min _{\widetilde \beta }}{\left\| {Y - \widetilde \phi \widetilde \beta } \right\|^2}$$
 
The normal equation is stable in the orthonormal basis:

$$\widetilde \beta  = {\left( {{{\widetilde \phi }^T}\widetilde \phi } \right)^{ - 1}}{\widetilde \phi ^T}Y$$
 
It is easy to see that  ${\widetilde \phi ^T}\widetilde \phi $ is a diagonal matrix with diagonal elements $1$ corresponding to significant eigenvalues, and $0$ corresponding to insignificant ones. With the convention  ${\left( {{{\widetilde \phi }^T}\widetilde \phi } \right)^{ - 1}} = {\widetilde \phi ^T}\widetilde \phi $ (invert the significant diagonal elements and zero the insignificant ones), we get:

 $$\widetilde \beta  = {D^{ - \frac{1}{2}}}{P^T}{\phi ^T}Y$$

(notice, ${D^{ - \frac{1}{2}}}$  zeroes the lines corresponding to insignificant eigenvalues so there is no need to left multiply by  ${\widetilde \phi ^T}\widetilde \phi $.)
Hence: $\widehat Y = \widetilde \phi \widetilde \beta  = \phi P{D^{ - 1}}{P^T}{\phi ^T}Y = \beta Y$  where ${D^{ - 1}} = {\left( {{D^{ - \frac{1}{2}}}} \right)^2}$  has diagonal elements inverse of the significant eigenvalues in  $D$, zero for the insignificant eigenvalues, and:

\[
\boxed{
    \beta  = P{D^{ - 1}}{P^T}{\phi ^T}Y
}
\]

\section{Tikhonov (ridge) regularization}

Classic regression works best with regularization, the most common classic form of which is ridge regression, also called Tikhonov regularization, which adds a penalty on the norm of $\beta$  in the objective cost.
 
 \begin{eqnarray*}
\beta  & = & \arg {\min _\beta }\left[ {{{\left\| {Y - \phi \beta } \right\|}^2} + {\lambda ^2}{{\left\| \beta  \right\|}^2}} \right] \\
& = & \arg {\min _\beta }\left[ {{{\left\| {Y - \left( {\phi P{D^{ - \frac{1}{2}}}} \right)\left( {{D^{\frac{1}{2}}}{P^T}\beta } \right)} \right\|}^2} + {\lambda ^2}{{\left\| \beta  \right\|}^2}} \right]\\
 & = & P{D^{ - \frac{1}{2}}}\arg {\min _\gamma }\left[ {{{\left\| {Y - \widetilde \phi \gamma } \right\|}^2} + {\lambda ^2}{{\left\| {P{D^{ - \frac{1}{2}}}\gamma } \right\|}^2}} \right] \\
 & = & P{D^{ - \frac{1}{2}}}\arg {\min _\gamma }\left[ {{{\left\| {Y - \widetilde \phi \gamma } \right\|}^2} + {\lambda ^2}{{\left\| {{D^{ - \frac{1}{2}}}\gamma } \right\|}^2}} \right]\\
 & = & P{D^{ - \frac{1}{2}}}{\left[ {{{\widetilde \phi }^T}\widetilde \phi  + {\lambda ^2}{D^{ - 1}}} \right]^{ - 1}}{\widetilde \phi ^T}Y \\
 & = & P{D^{ - \frac{1}{2}}}{\left[ {{{\widetilde \phi }^T}\widetilde \phi  + {\lambda ^2}{D^{ - 1}}} \right]^{ - 1}}{D^{ - \frac{1}{2}}}{P^T}{\phi ^T}Y \\ 
 & = & P{\Lambda ^{ - 1}}{P^T}{\phi ^T}Y
\end{eqnarray*}

where ${\Lambda ^{ - 1}}$  has diagonal elements $\frac{1}{{{D_{ii}} + {\lambda ^2}}}$  where ${D_{jj}}$  is significant, zero otherwise. And we get:
 
\[
\boxed{
    \beta \left( \lambda  \right) = P\Lambda {\left( \lambda  \right)^{ - 1}}{P^T}{\phi ^T}Y
}
\]

The Tikhonov parameter $\lambda $  can be found e.g. by cross validation: 

$$\lambda  = \arg {\min _\lambda }{\left\| {{Y_V} - {\phi _V}\beta \left( \lambda  \right)} \right\|^2}$$
 
where  ${\phi _V} = \phi \left( {{X_V}} \right)$, $\left( {{X_V},{Y_V}} \right)$  form a validation set of independent examples and  $\beta \left( \lambda  \right)$ is the result of a ridge regression over the training set with Tikhonov parameter  $\lambda $, obtained with the boxed formula above. The objective function can be expanded:

\[
\begin{array}{lll}
f\left( \lambda  \right) & = & {\left\| {{Y_V} - {\phi _V}\beta \left( \lambda  \right)} \right\|^2}\\
 & = & {\left( {{Y_V} - {\phi _V}\beta \left( \lambda  \right)} \right)^T}\left( {{Y_V} - {\phi _V}\beta \left( \lambda  \right)} \right)\\
 & = & {\left( {{Y_V} - {\phi _V}P\Lambda {{\left( \lambda  \right)}^{ - 1}}{P^T}{\phi ^T}Y} \right)^T}\left( {{Y_V} - {\phi _V}P\Lambda {{\left( \lambda  \right)}^{ - 1}}{P^T}{\phi ^T}Y} \right)\\
 & = & \left( {{Y_V}^T - {Y^T}\phi P\Lambda {{\left( \lambda  \right)}^{ - 1}}{P^T}{\phi _V}^T} \right)\left( {{Y_V} - {\phi _V}P\Lambda {{\left( \lambda  \right)}^{ - 1}}{P^T}{\phi ^T}Y} \right)\\
 & = & {Y_V}^T{Y_V} - {Y^T}\phi P\Lambda {\left( \lambda  \right)^{ - 1}}{P^T}{\phi _V}^T{Y_V} - {Y_V}^T{\phi _V}P\Lambda {\left( \lambda  \right)^{ - 1}}{P^T}{\phi ^T}Y\\
 &  & + {Y^T}\phi P\Lambda {\left( \lambda  \right)^{ - 1}}{P^T}{\phi _V}^T{\phi _V}P\Lambda {\left( \lambda  \right)^{ - 1}}{P^T}{\phi ^T}Y
\end{array}
\]
 
Since ${Y_V}^T{Y_V}$  doesn’t depend on  $\lambda $, we minimize:
 
 \[\begin{array}{lll}
g\left( \lambda  \right) & = & {Y^T}\phi P\Lambda {\left( \lambda  \right)^{ - 1}}{P^T}{\phi _V}^T{\phi _V}P\Lambda {\left( \lambda  \right)^{ - 1}}{P^T}{\phi ^T}Y\\
 & - & {Y^T}\phi P\Lambda {\left( \lambda  \right)^{ - 1}}{P^T}{\phi _V}^T{Y_V}\\
 & - & {Y_V}^T{\phi _V}P\Lambda {\left( \lambda  \right)^{ - 1}}{P^T}{\phi ^T}Y\\
 & = & {K^T}\Lambda {\left( \lambda  \right)^{ - 1}}M\Lambda {\left( \lambda  \right)^{ - 1}}K - {K^T}{\left( \lambda  \right)^{ - 1}}L -  {L^T}\Lambda {\left( \lambda  \right)^{ - 1}}K\\
 & = & {K^T}\Lambda {\left( \lambda  \right)^{ - 1}}M\Lambda {\left( \lambda  \right)^{ - 1}}K - 2{K^T}{\left( \lambda  \right)^{ - 1}}L\\
 & = & {K^T}\Lambda {\left( \lambda  \right)^{ - 1}}\left[ {M\Lambda {{\left( \lambda  \right)}^{ - 1}}K - 2L} \right]
\end{array}\]
 
where  \[K = {P^T}{\phi ^T}Y\left\{ {n \times 1} \right\},L = {P^T}{\phi _V}^T{Y_V}\left\{ {n \times 1} \right\} \text{ and } M = {P^T}{\phi _V}^T{\phi _V}P\left\{ {n \times n} \right\}\]

Optimization may be efficiently performed by a classic one-dimensional minimization procedure.

\section{Differential Regression}

In addition to inputs  $X$  and labels  $Y$, we have derivatives labels  $Z$ whose columns ${Z_j}$  are the differentials of $Y$  to  $X_j$. Denote ${\phi _j}$  the matrix of derivatives of the basis functions $\phi $  wrt ${X_j}$. Linear regression makes value predictions $ \widehat Y = \phi \beta $  and derivatives predictions  ${\widehat Z_j} = {\phi _j}\beta $. We now minimize a cost combining value and derivatives errors:
 
 $$\beta  = \arg {\min _\beta }\left[ {{{\left\| {Y - \phi \beta } \right\|}^2} + \sum\limits_j {{\lambda _j}{{\left\| {{Z_j} - {\phi _j}\beta } \right\|}^2}} } \right]$$
 
where  ${\lambda _j} = {\lambda ^2}\frac{{{{\left\| Y \right\|}^2}}}{{{{\left\| {{Z_j}} \right\|}^2}}}$ (norms are computed across examples) ensures that the components of the cost are of similar magnitudes. The hyperparameter  $\lambda $ has little effect and generally left to $1$.

It is not hard to see that this minimization is analytically solved with the adjusted normal equation:
 
$$\beta  = {\left( {{\phi ^T}\phi  + \sum\limits_j {{\lambda _j}{\phi _j}^T{\phi _j}} } \right)^{ - 1}}\left( {{\phi ^T}Y + \sum\limits_j {{\lambda _j}{\phi _j}^T{Z_j}} } \right)$$ 
 
This is, again, a theoretical equation, unstable in practice. As before, we change basis by eigenvalue decomposition of:
 
 $${\phi ^T}\phi  + \sum\limits_j {{\lambda _j}{\phi _j}^T{\phi _j}}  = PD{P^T}$$
 
-- beware, notations have changed so $P$  and $D$  denote different (respectively unitary and diagonal) matrices than before.
Changing basis as before:  $\widetilde \phi  = \phi P{D^{ - \frac{1}{2}}}$ (where, as previously, ${D^{ - \frac{1}{2}}}$  has zero diagonal elements where the eigenvalues in  $D$ are insignificant) we notice that:
 
 $${\widetilde \phi _j} \equiv \frac{{\partial \widetilde \phi }}{{\partial {X_j}}} = \frac{{\partial \left( {\phi P{D^{ - \frac{1}{2}}}} \right)}}{{\partial {X_j}}} = \frac{{\partial \phi }}{{\partial {X_j}}}P{D^{ - \frac{1}{2}}} = {\phi _j}P{D^{ - \frac{1}{2}}}$$
 
Performing the minimization in the ‘tilde’ basis:
 
 $$\widetilde \beta  = \arg {\min _{\widetilde \beta }}\left[ {{{\left\| {Y - \widetilde \phi \widetilde \beta } \right\|}^2} + \sum\limits_j {{\lambda _j}{{\left\| {{Z_j} - {{\widetilde \phi }_j}\widetilde \beta } \right\|}^2}} } \right]$$
 
We have the normal equation:
 
 \[
 \begin{array}{lll}
\widetilde \beta  & = & {\left( {{{\widetilde \phi }^T}\widetilde \phi  + \sum\limits_j {{\lambda _j}{{\widetilde \phi }_j}^T{{\widetilde \phi }_j}} } \right)^{ - 1}}\left( {{{\widetilde \phi }^T}Y + \sum\limits_j {{\lambda _j}{{\widetilde \phi }_j}^T{Z_j}} } \right)\\
 & = & {\left[ {\left( {{D^{ - \frac{1}{2}}}{P^T}} \right)\left( {{\phi ^T}\phi  + \sum\limits_j {{\lambda _j}{\phi _j}^T{\phi _j}} } \right)\left( {P{D^{ - \frac{1}{2}}}} \right)} \right]^{ - 1}}\left[ {\left( {{D^{ - \frac{1}{2}}}{P^T}} \right)\left( {{\phi ^T}Y + \sum\limits_j {{\lambda _j}{\phi _j}^T{Z_j}} } \right)} \right]\\
 & = & {D^{ - \frac{1}{2}}}{P^T}\left( {{\phi ^T}Y + \sum\limits_j {{\lambda _j}{\phi _j}^T{Z_j}} } \right)
\end{array}
\]
 
Predicted values are given by: 
 
 \[\widehat Y = \widetilde \phi \widetilde \beta  = \phi P{D^{ - \frac{1}{2}}}{D^{ - \frac{1}{2}}}{P^T}\left( {{\phi ^T}Y + \sum\limits_j {{\lambda _j}{\phi _j}^T{Z_j}} } \right) = \phi P{D^{ - 1}}{P^T}\left( {{\phi ^T}Y + \sum\limits_j {{\lambda _j}{\phi _j}^T{Z_j}} } \right) = \phi \beta \]
 
where ${D^{ - 1}} = {\left( {{D^{ - \frac{1}{2}}}} \right)^2}$  is defined as previously (with zeroes on insignificant eigenvalues) and:

\[
\boxed{
    \beta  = P{D^{ - 1}}{P^T}\left( {{\phi ^T}Y + \sum\limits_j {{\lambda _j}{\phi _j}^T{Z_j}} } \right)
}
\]

Note that this is all consistent, in particular, derivatives predictions are given by:
 
 \[{\widehat Z_j} = {\phi _j}\beta  = {\phi _j}P{D^{ - 1}}{P^T}\left( {{\phi ^T}Y + \sum\limits_j {{\lambda _j}{\phi _j}^T{Z_j}} } \right) = \left( {{\phi _j}P{D^{ - \frac{1}{2}}}} \right)\left[ {{D^{ - \frac{1}{2}}}{P^T}\left( {{\phi ^T}Y + \sum\limits_j {{\lambda _j}{\phi _j}^T{Z_j}} } \right)} \right] = {\widetilde \phi _j}\widetilde \beta \]

\section*{Conclusion}

We derived a normal equation (SVD style) for differential regression (in the sense of the working paper's differential machine learning) and verified its effectiveness in a public demonstration notebook. Differential regularization vastly outperforms classic variants, including ridge, and without consuming additional data or needing any form of additional regularization or cross validation. Just like Tikhonov regularization, differential regularization is analytic and extremely effective, as seen in the demonstration notebook. Unlike Tikhonov, differential regularization is unbiased, as demonstrated in another appendix.

\pagebreak 

\chapter{Supervised Learning without Supervision: \texorpdfstring{\\}{}
    Wide and Deep Architecture\texorpdfstring{\\}{}
    and Asymptotic Control}
\counterwithout{section}{chapter}
\setcounter{section}{0}
\label{app4}
\section*{Introduction}

Modern deep learning is very effective at function approximation, especially in the differential form presented in the working paper. But training of neural networks is a nonconvex problem, without guaranteed convergence  to the global minimum of the cost function or close\footnote{It is also prone to overfitting, so generalization is not guaranteed even with minimum MSE on the training set. Differential machine learning considerably helps, as abundantly coommented in the working paper and other appendices.}. Neural networks are usually trained under close human supervision and it is hard to execute it reliably behind the scenes. 

This is of particular concern in Derivatives risk management, where automated procedures cannot be implemented in production without strong guarantees. Vast empirical evidence, that modern training heuristics (data normalization, Xavier-Glorot initialization, ADAM optimization, one-cycle learning rate schedule...) often combine to converge to acceptable minima, is not enough. Risk management is not built on faith but on mathematical guarantees.

In this appendix, we see how and to what extent guarantees can be established for training neural networks with a special architecture called 'wide and deep', also promoted by Google in the context of recommender systems. We show that wide and deep learning is \emph{guaranteed} to do at least as well as classic regression, opening the possibility of training without supervision.

We also discuss asymptotic control, another key requirement for reliable implementation in production.

\section{Wide and Deep Learning}

\subsection{Wide vs Deep}

\subsubsection{Wide regression}

Classic regression (which we call \emph{wide learning} for reasons apparent soon) finds an approximation $\hat f$ of a target function $f: \mathbb{R}^n \to \mathbb{R}$ as a linear combination of a predefined set of $p$ basis functions $\phi_j$ of inputs $x$ in dimension $n$:

$$
    \hat f \left( x ; w \right) = \sum_{j=1}^{p} w_j \phi_j \left( x \right) = \phi \left( x \right) w
$$

\noindent by projection onto the space of functions spanned by the basis functions $\phi_j$. With a training set of $m$ examples given by the matrix $X$ of shape $m \times n$, with labels stacked in a vector $Y$ of dimension $m$, the $p$ learnable weights $w_j$ are estimated by minimization of the mean squared error (MSE), itself an unbiased estimation of the distance $||\hat f -f||^2$ in $L^2$:

$$
    \hat w = argmin_w MSE = \sum_{i=1}^m \left[ \phi \left( X ^{\left( i \right)} \right) w - Y ^{\left( i \right)} \right]^2
$$

It is immediately visible that the objective $MSE$ is convex in the weights $w$. The optimization is well defined with a unique minimum, easily found by canceling the gradient of the $MSE$ wrt $w$, resulting in the well known \emph{normal equation}:

$$
    \hat w = \left( \Phi^T \Phi \right)^{-1} \Phi^T Y
$$

\noindent where $\Phi$ is the $m \times p$ matrix stacking basis functions of inputs in its row vectors:

$$
\Phi  ^{\left( i \right)} = \phi \left( X ^{\left( i \right)}\right)
$$

Let us call \emph{input dimension} the dimension $n$ of $x$ and \emph{regression dimension} the dimension $p$ of $\phi$. In low (regression) dimension, the normal equation is tractable but subject to numerical trouble when the matrix $\Phi^T \Phi$ is near singular. This is resolved by SVD regression (see \ref{app3}), a safe implementation of the projection operator so the problem is still convex and analytically solvable. In high dimension, the inversion or SVD decomposition may become intractable, in which case the argmin of the MSE is found numerically, e.g. by a variant of gradient descent. Importantly, the problem remains convex so numerical optimizations like gradient descent are guaranteed to converge to the unique minimum (modulo appropriate learning rate schedule).

This is all good, but it should be clear that the practical performance of classic regression is highly dependent on the relevance of the basis functions $\phi_j$ for the approximation of the true function $f$, mathematically measured by the $L^2$ distance between the true function and the space spanned by the basis functions, of which the minimum MSE is an estimate. 

One strategy is pick a vast number of basis functions $\phi_j$ so that their combinations approximate \emph{all} functions to acceptable accuracy. For example, the set of all \emph{monomials} of $x$ of the form:

$$
    \phi_j \left( x\right) = \prod _{j=1}^n x_j^{k_j} \text{ such that }  \sum_{j=1}^n k_j \leq K
$$

\noindent are dense in $L^2 \left(\mathbb{R}^n \right)$ so polynomial regression has the \emph{universal approximation} property: it approximates all functions to arbitrary accuracy by growing degree $K$. Regardless, this strategy is almost never viable in practice due to the \emph{course of dimensionality}. Readers may convince themselves that the number of monomials of degree up to $K$ in dimension $n$ is:

$$
    \frac {\left( n + K \right)!} {n! K!}
$$

\emph and grows exponentially in the input dimension $n$ and polynomial degree $K$. A cubic regression in dimension $20$ has $1,771$ monomials. A degree $7$ polynomial regression has $888,030$. Given exponentially growing number of learnable parameters $w$ (same as number of basis functions), the size $m$ of the dataset must grow \emph{at least as fast} for the problem to stay well defined. In most contexts of practical relevance, dimension of this magnitude is both computationally intractable and bound to overfit training noise, even when dimension $n$ was previously reduced with a meaningful method like differential PCA (see \ref{app2}). The same arguments apply to all other bases of functions besides polynomials: Fourier harmonics, radial kernels, cubic splines etc. They are all affected by the same curse and only viable in low dimension. 

Regression is only viable in practice when basis functions are carefully selected with handcrafted rules from contextual information. One example is the classic Longstaff-Schwartz algorithm (LSM) of 2001, originally designed for the regression of the continuation value of Bermudan options in the Libor Market Model (LMM) of Musiela and al. (1995). The Markov state of LMM is high dimensional and includes all forward Libor rates up to a final maturity, e.g. with 3m Libors up to maturity 30y, dimension is $n=120$. To regress the value of a Bermudan swaption in such high dimension is hopeless. Instead, classic implementations regress on low dimensional features (i.e. a small number of nonlinear functions) of the state, called \emph{regression variables}. It is known that Bermudan options on call dates are mainly sensitive to the swap rates to maturity and to the next call (assuming deterministic volatility and basis). Instead of attempting regression in dimension $120$, effective implementations of LSM simply regress on those two functions of the state, effectively reducing regression dimension from $120$ to $2$.

This is very effective, but it takes prior knowledge of the generative model. We can safely apply this methodology because we know that this is a Bermudan option in a model with deterministic volatility and basis. Careful prior study determined that the value mainly depends on two regression variables, the two swap rates, so we could hardcode the transformation of the $120$ dimensional state into a $2$ dimensional regression vector as part of LSM implementation. This all fails when the transaction is not a standard Bermudan swaption, or with a different simulation model (say, with stochastic volatility, when volatility state is another key regression variable). The methodology cannot be applied to arbitrary schedules of cash flows, simulated in arbitrary models. In practice, \emph{regression cannot learn from data alone}. This is why it doesn't qualify as \emph{artificial intelligence} (AI). Regression is merely a fitting procedure. Intelligence lies in the selection of basis functions, which is performed by hand and hardcoded as a set of rules. In fact, the whole thing can be seen as a neural network (NN) with fixed, nonlearnable hidden layers, as shown in \figref{RegAsNN}.

\begin{figure}[htp]
\centering
\includegraphics[scale=0.6]{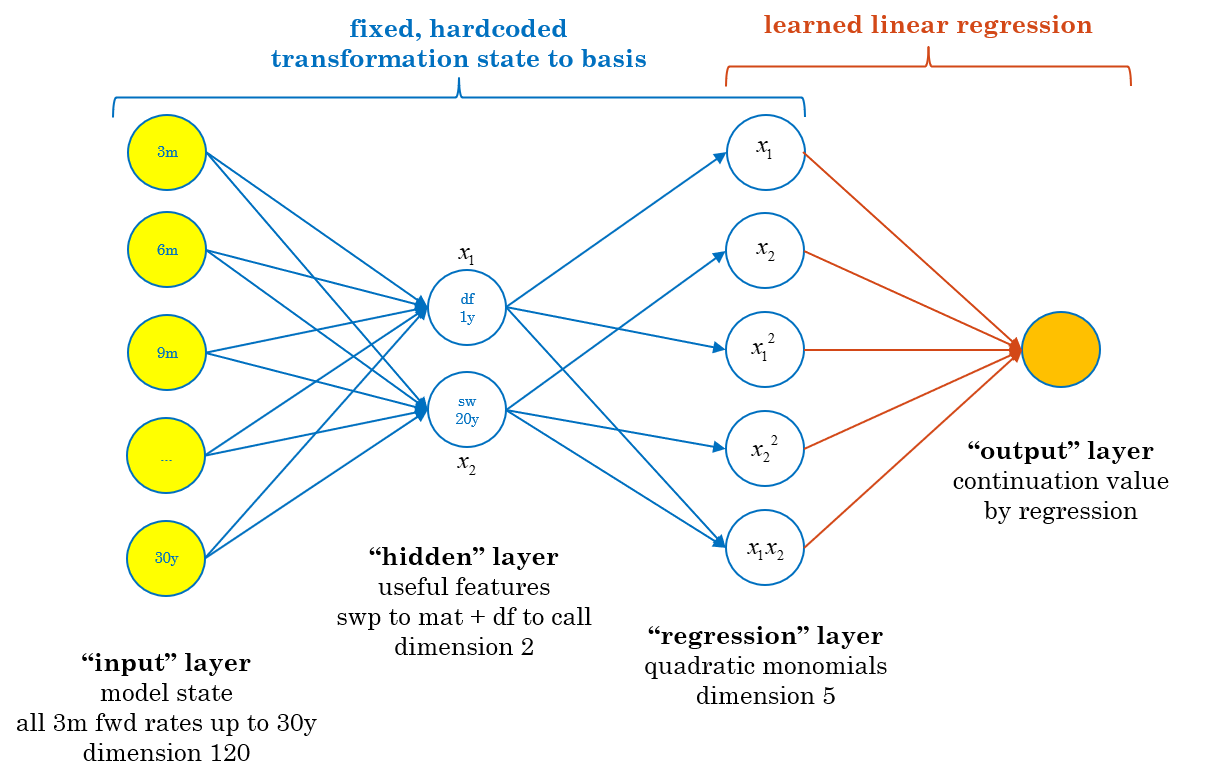}
\caption{LSM regression as NN with fixed hidden layers}
\label{fig:RegAsNN}
\end{figure}

\subsubsection{Deep neural networks}

By contrast, neural networks are 'intelligent' constructs, capable of learning from data alone. NN are extensions of classic regression. In fact, they are identical to regression safe for one crucial difference: NN internalize the selection of basis functions and \emph{learn them from data}. For example, consider a NN with $4$ hidden layers of $20$ softplus activated units. The output layer is a classic regression over the basis functions identified in the \emph{regression layer} (the last hidden layer). When the NN is trained by minimization of the MSE, the hidden weights learn to encode the $20$ 'best' basis functions  among the (very considerable) space of functions attainable with the deep architecture. Optimization finds the best basis functions in the sense of the MSE, which itself approximates the distance between the true function and the space spanned by the basis functions. Hence, training a neural network really boils down to finding the appropriate, low dimensional regression space, often called \emph{feature extraction} in machine learning (ML). 

The strong similarity of NN to regression is illustrated on \figref{NN} where we also see the one major difference: hidden layers are no longer fixed, they have learnable connection weights.

\begin{figure}[htp]
\centering
\includegraphics[scale=0.6]{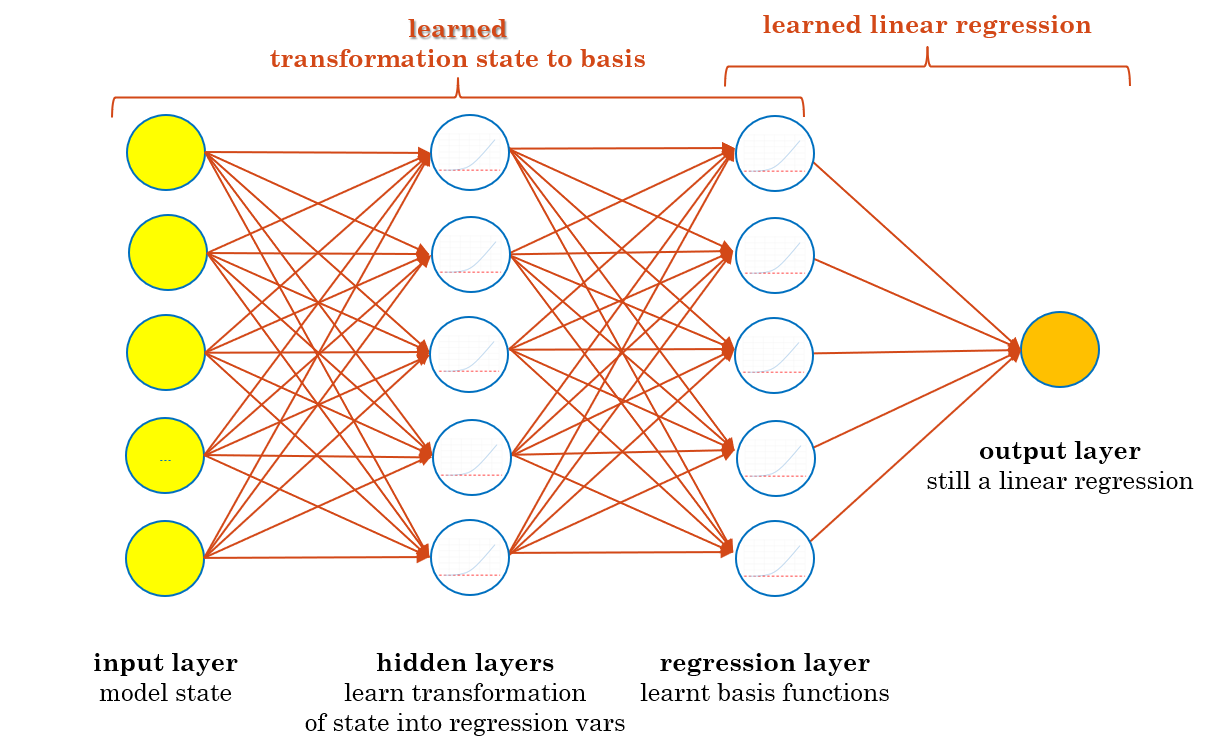}
\caption{NN with two learnable hidden layers}
\label{fig:NN}
\end{figure}

This is what makes NN so powerful, e.g. for solving high dimensional problems in computer vision or natural language processing. NN cruise through the curse of dimensionality be \emph{learning} the limited dimension space that best approximates the target function. They learn from data alone without handcrafted rules based on prior knowledge or specific to a given context. In particular, NN effectively approximate high dimensional functions from data alone,  in finance or elsewhere, and they may even outperform regression on handcrafted features, i.e. find better basis functions from data than those extracted by hand from prior knowledge.

In return, training a neural network is famously \emph{not} a convex problem. NN generally include many learnable connection weights ($1224 + 20n$ with $4$ hidden layers of $20$ units) and the MSE function of the weights has been shown to be of complicated topology, including multiple local minima and saddle points. There famously exists no algorithm guaranteed to find the global minimum in finite time. Despite extremely active research resulting in powerful \emph{heuristic} improvements, training NN remains an art as much as a science. NN are often trained by hand over long periods where engineers slowly tweak architecture and hyperparameters until they obtain the desired behaviour in a given context. Modern training algorithms 'generally' converge to 'acceptable' minima, but such terms don't cut ice in mathematics, and they shouldn't in risk management either. 

In order to automate training and implement its automatic execution, behind the scenes and \emph{without human supervision}, we need sufficient mathematical guarantees. While it may look at first sight as a hopeless endeavour, we will see that the analysis performed in this paragraph allows to combine NN with regression in a meaningful and effective manner to establish important worst case guarantees.

\subsection{Wide \emph{and} deep}

\subsubsection{Mixed architecture}

While regression is often opposed to deep learning in literature, a natural approach is to combine their benefits, by regression on both learnable \emph{deep} units and fixed \emph{wide} units, as illustrated in \figref{DeepWide}. Mathematically, the ouput layer is a linear regression on the concatenation of the deep layer $z_{L-1}$ (the last hidden layer of the deep network) and a set of fixed basis function $\phi$ (a.k.a. the wide layer):

$$
    \hat f \left( x ; w \right) = z_{L-1} \left( x ; w_{hidden} \right) w_{deep} + \phi \left( x \right) w_{wide}
$$

\noindent and it immediately follows that the differentials of predictions wrt inputs are:

$$
    \frac {\partial \hat f \left( x ; w \right)} {\partial x_j} = \frac{ \partial z_{L-1} \left( x ; w_{hidden} \right) } {\partial x_j } w_{deep} + \frac {\partial \phi \left( x \right)} {\partial x_j} w_{wide}
$$

\noindent where ${ \partial  z_{L-1}  } / {\partial x }$ are differentials of the deep network computed by backpropagation (actually a platform like TensorFlow can compute the whole gradient of the wide and deep net behind the scenes) and ${\partial \phi } / {\partial x}$ are the known derivatives of the fixed basis functions. Hence, the architecture is trivially implemented by:

\begin{enumerate}
    
    \item Add a classic regression term $\phi \left( x \right) w_{wide}$ to the output of the deep neural network.
    
    \item Adjust the gradients of output wrt inputs (given by backpropagation through the deep network) by $\left({\partial \phi \left( x \right)} / {\partial x_j} \right) w_{wide}$ (or leave it to TensorFlow).

\end{enumerate}

An implementation in code is given in Geron's textbook, second ed. chapter 10. 

\begin{figure}[htp]
\centering
\includegraphics[scale=0.6]{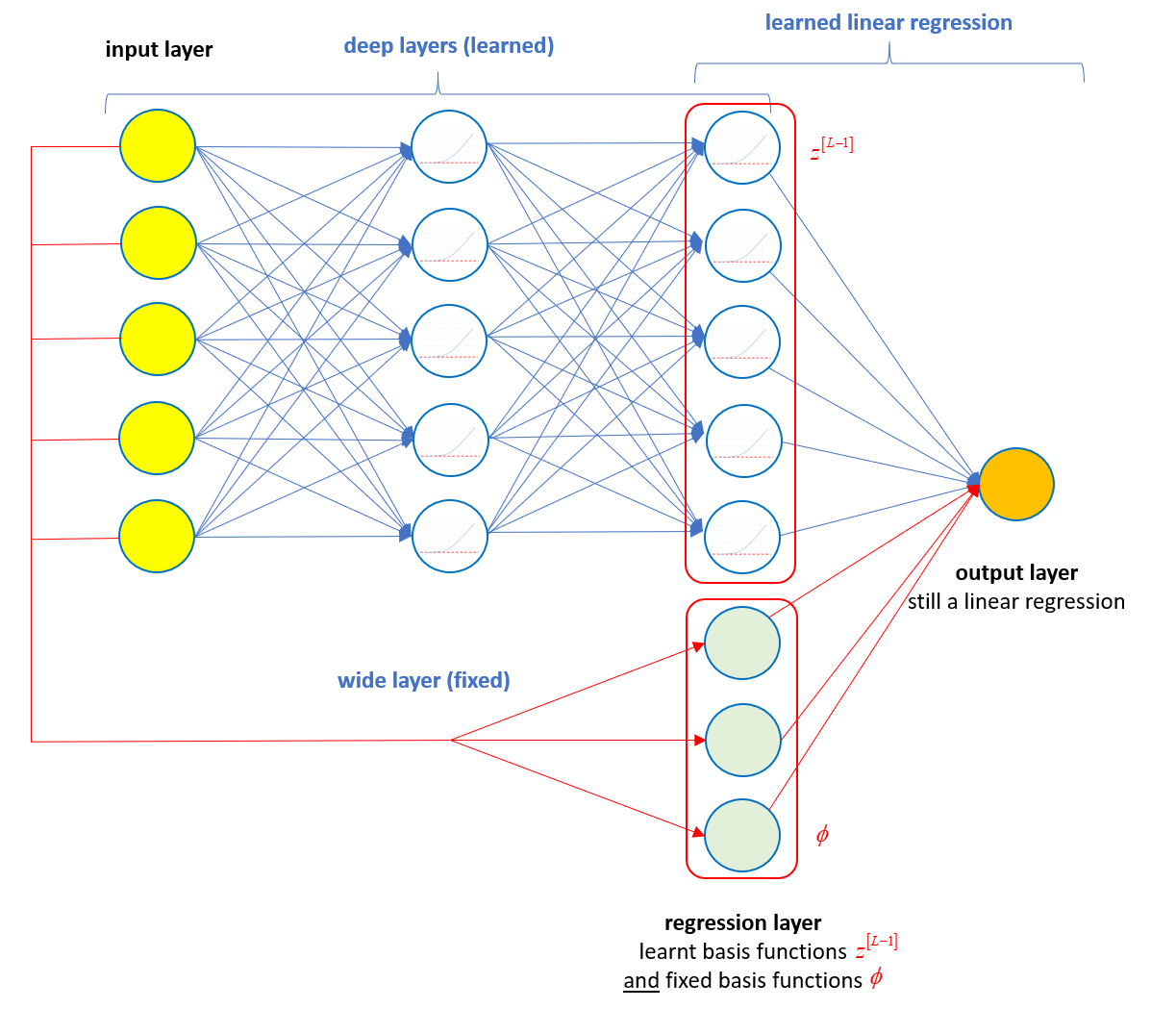}
\caption{Wide and deep architecture}
\label{fig:DeepWide}
\end{figure}

The concatenation of the wide and deep layer doesn't have to be the final regression layer. The architecture is flexible and supports all kinds of generalization. For example, we could insert a couple of fully connected layers \emph{after} the wide and deep layer to discover meaningful nonlinear combinations of wide and deep basis functions in the final regression layer. Backpropagation equations are easily updated (or left to automatic differentiation in TensorFlow). 

The idea is natural, and certainly not new. It was popularized by Google under the name "wide and deep learning", in the context of recommender systems (\url{https://arxiv.org/abs/1606.07792}), although from a different perspective. Specifying a number of fixed regression functions in the wide layer should help training by restricting search for additional basis functions in the deep layers to dissimilar functions. For example, when the wide layer is a copy of the input layer $x$ ($\phi = id$), it handles all linear functions of $x$ and specializes the deep layers to a search for nonlinear functions (since another linear function in the deep regression layer would not help reduce MSE). In other terms, Google presented the wide and deep architecture as a training improvement, and it may well be that it does improve performance significantly with very deep, complex architectures. In our experience, the improvement is marginal with the simple architecture sufficient for pricing function approximation, but the wide and deep architecture still has a major role to play, because it provides guarantees and allows a safe implementation of automated training without supervision.

\subsubsection{Worst case convergence guarantee}

It is general wisdom that minimization of the MSE with NN doesn't offer any sort of guarantee. This is not entirely correct, though. Consider the MSE as a function of the connection weights of the output layer alone. This is evidently a convex function. In fact, since the output layer is exactly a linear regression on the regression layer, the optimal weights are even given in closed form by the normal equation:

$$
    \hat w_\text{output} = \left( {z_{L-1}}^T z_{L-1}\right)^{-1} {z_{L-1}}^T Y
$$

\noindent or its SVD equivalent (see \ref{app3}). Recall, while numerical optimization may not find the global minimum, it is always guaranteed to converge to a point with uniform zero gradient. In particular, training converges to a point where the derivatives of the MSE to the \emph{output} connection weights are zero. And since the MSE is convex in \emph{those} weights, \emph{the projection onto the basis space is always optimal.} Training may converge to 'bad' basis functions, but the approximation \emph{in terms of these basis functions} is always as good as it can be. It immediately follows that, with a deep and wide architecture, we have a meaningful worst case guarantee: the approximation is least as good as a linear regression on the wide units. In practice, we get an orders of magnitude better performance from the deep layers, but it is the worst case guarantee that gives us permission to train without supervision. In practice, convergence may be checked by measuring the norm of the gradient, or, optimization may be followed by an analytic implementation of the normal equation wrt the combined regression layer (ideally in the SVD form of \ref{app3}).

\subsubsection{Selection of wide basis}

Of course, the worst case guarantee is only as good as the choice of the wide functions. An obvious choice is a straightforward copy of the input layer. The wide layer handles all linear functions of the inputs, hence the worst case result is a linear regression. Another strategy is also add the squares of the input layers, and perhaps the cubes, depending on dimension, but not the cross monomials, which would bring back the curse of dimensionality.

A much more powerful wide layer may be constructed in combination with differential PCA (see \ref{app2}), which reduces the dimension of inputs and orders them by relevance, in a basis where differentials are orthogonal. This means that the input column $X_1$ affects targets most, followed by $X_2$ etc. Because inputs are presented in a relevant hierarchy, we may build a meaningful wide layer with a richer set of basis functions applied to the most relevant inputs. For example, we could use all monomials up to degree $3$ on the first two inputs ($10$ basis functions), monomials of degree less than two on the next three inputs (another nine basis functions), and the other $n-5$ inputs raised to power $1$, $2$ and maybe $3$ (up to $3n-15$ additional functions). Because of the  differential PCA mechanism, a plain regression on these basis functions bears acceptable results by itself, especially with differential regression (see \ref{app3}), and this is only the \emph{worst case} guarantee, with orders of magnitude better average performance.

All those methods learn from data alone, with worst case guarantees. In cases where meaningful basis functions are handcrafted from contextual information and reliable hardcoded rules, like for Bermudan options in LMM with LSM, \emph{wide and deep networks still outperform} as we see next. 

\subsubsection{Outperformance: $w+d \geq w$ w and $w+d \geq d$  }

If follows from what precedes that the wide and deep architecture is guaranteed to find a better fit than either the deep network or the wide regression alone.

In particular, wide and deep networks outperform classic regression, even on relevant handcrafted basis functions. Not only are they guaranteed to fit training data at least as well, they will also often find meaningful features missing from the wide basis. For example, Bermudan options are \emph{mainly} sensitive to rates to expiry and next call, but the shape of the yield curve also matters to an extent. The deep layers should identify the additional relevant factors during training. Finally, wide and deep nets are resilient to change. Add stochastic volatility in the LMM, regression no longer works without modification of the code to account for additional basis functions including volatility state. Wide and deep nets would work without modification, building volatility dependent features in their deep layers. 

\section{Asymptotic control}

\subsection{Elementary asymptotic control}

\subsubsection{Enforce linear asymptotics}

Another important consideration for unsupervised training is the performance of the trained approximation on \emph{asymptotics}. This is particularly crucial for risk management applications like value at risk (VAR), expected loss (EL) or FRTB, which focuses on the behaviour of trading books in extreme scenarios. Asymptotics are hard because they are generally learned from little to no data in edge scenarios. In other terms, the asymptotic behaviour of the approximation is an \emph{extrapolation} problem and reliable extrapolation is always harder, for instance, polynomial regression absolutely cannot be trusted.

As always, we want to control asymptotics from data alone and not explore methods based on prior knowledge of the correct asymptotics. For instance, a European call is known to have flat left asymptotic and linear right asymptotic with slope $1$. If we know that the transaction is a European call, the correct asymptotics could be enforced by a variety of methods, see e.g. Antonov and Piterbarg for cutting edge. But that only works when we know for a fact that we are approximating the value of a European call. What we want is a general algorithm without applicable without other knowledge than a simulated dataset. 

In finance, linear asymptotics are generally considered fair game for pricing functions, with an unknown slope to be estimated from data. For instance, it is common practice to enforce a zero second derivative boundary condition when pricing with finite difference methods (FDM)\footnote{Although overreliance on this common assumption may be dangerous: the Derivatives industry lost billions in 2008 on variance swaps and CMS caps, precisely due to nonlinear asymptotics.}. Linear asymptotics are guaranteed for neural networks as long as the activations are asymptotically linear. This is the case e.g. for common RELU, ELU, SELU or softplus activations, but not sigmoid or tanh, which asymptotics are flat, hence, to be avoided for pricing approximation\footnote{Recall that \emph{differential} deep learning requires $C^1$ activation, ruling out RELU and SELU and leaving only the very similar ELU or softplus among common activations.}. 

\figref{asymptotics} compares the asymptotics of polynomial and neural approximations for a call price in Bachelier's normal model, obtained with our demonstration notebooks \emph {DifferentialML.ipynb} and \emph{DifferentialRegression.ipynb} on \url{https://github.com/differential-machine-learning/notebooks} (dimension 1, 8192 training examples). The trained approximation is voluntarily tested on an unreasonably wide range of inputs in order to highlight asymptotic behaviour. Unsurprisingly, polynomial regression terribly misbehaves whereas neural approximation fares a lot better due to linear extrapolation. The comparison is of course unfair. The outperformance of the neural net in edge scenarios is only due to linear asymptotics, which can be equally enforced for linear regression\footnote{By cropping basis functions with linear extrapolation outside of a given domain.}. The point here is that we want to enforce linear asymptotics\footnote{While maintaining awareness that not all transactions are linear on edges.}, something given with neural networks, and doable with some effort with regression.

\begin{figure}[htp]
\centering
\includegraphics[scale=0.6]{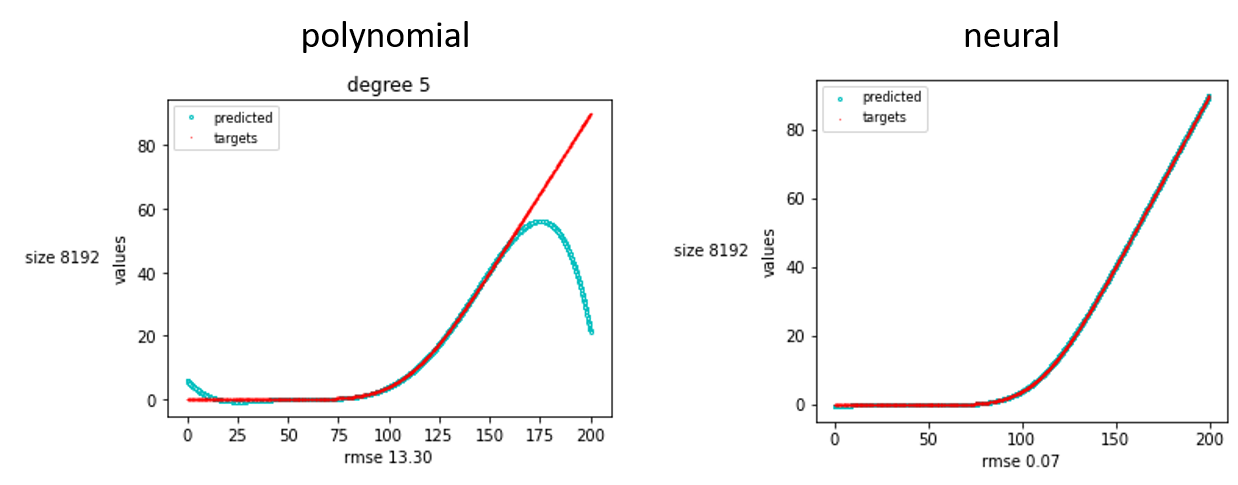}
\caption{Asymptotics in a polynomial and neural approximation of a European call price}
\label{fig:asymptotics}
\end{figure}

\subsubsection{Oversample extreme scenarios}

Enforcing linear asymptotics is not enough, we must also learn the slope from data. What makes it difficult is the typical sparsity of data on the edges of the training domain, especially when training data is produced with Monte-Carlo, and noisy labels e.g. from LSM simulations certainly don't help.

By far, the easiest walkaround is to simulate a larger number of edge examples. Recall, we may sample training inputs in any way we want. It is only the labels that must be computed in complete agreement with the pricing model, either by conditional sampling (sample labels) or conditional expectation (ground truth labels). Hence, we sample training examples over a domain and with a distribution reflecting the intended use of the trained approximation. In applications where asymptotics are important, we want many training examples in edge scenarios.

When training examples are sampled with Monte-Carlo simulations, it is particularly simple to oversample extreme scenarios \emph{by increasing volatility from today to the horizon date} in the generative simulations. We implemented this simple method in our demonstration notebooks. As expected, increasing volatility to horizon date effectively resolves asymptotic behaviour, as illustrated on \figref{asymptoticsFixed} for degree 5 polynomial regression of the Bachelier call price.

\begin{figure}[htp]
\centering
\includegraphics[scale=0.6]{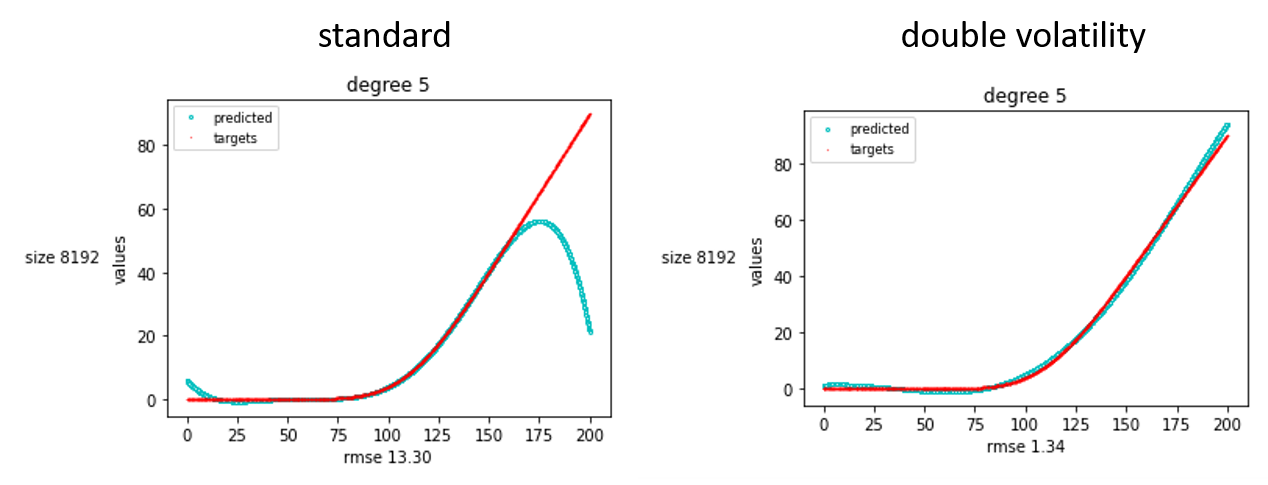}
\caption{Fixing polynomial asymptotics with increased volatility}
\label{fig:asymptoticsFixed}
\end{figure}

Note that increased volatility fixes asymptotics but deteriorates accuracy in the interior domain. By enforcing more examples on edge scenarios, we learn from fewer examples elsewhere, resulting in a loss of quality. 

Further, it is fair game to increase volatility or change model parameters in any way \emph{before horizon}, but the (conditional) simulation \emph{after horizon date} must exactly follow the pricing model or we get biased labels. With multiple horizon dates, the simple walkaround no longer works: between two horizon dates, we cannot simultaneously simulate an increased volatility for the state variables, and the original volatility for the cashflows\footnote{One solution is to simulate Monte-Carlo paths with a form of \emph{importance sampling} where samples are normalized by a likelihood ratio to compensate for increased volatility. However, importance sampling may raise numerical problems, since likelihood ratios between measures with different volatilities diverge in the continuous limit.}.  

For these reasons, our simple walkaround is certainly not an optimal solution, but it is extremely simple to comprehend and implement, with reasonable performance for such a trivial method. The more advanced algorithms introduced next perform a lot better, but with significant implementation effort.

\subsection{Advanced asymptotic control}

\subsubsection{Ground truth labels in edge examples}

In the main article and \ref{app1}, we have opposed \emph{ground truth learning}, where labels are numerically computed conditional expectations, to \emph{sample learning} where labels are samples drawn from the conditional distribution, e.g. by simulation of one Monte-Carlo path. We concluded that ground truth learning is not viable in many contexts of practical relevance because of the computational cost of conditional expectations, and that sample learning offers a viable, consistent alternative.

The opposition between the two doesn't have to be black and white. In fact, many intermediate solutions exist. Ground truth labels are computed by averaging a large number of samples, theoretically infinity. Sample labels are (averages of) one sample each. We could as well compute labels by averaging an intermediate number $N$ of samples, reducing variance by $\sqrt{N}$ in return for a computation cost linearly increasing in $N$. Notice that in the demonstration notebooks, we computed labels by averaging \emph{two} antithetic paths. As a result, the variance of the noise is reduced by a factor two, but we can simulate half as many examples for a fixed computation load. Hence, benefits balance out but we get the additional benefit of antithetic sampling, making it worthwhile.

This realization leads to a powerful asymptotic control algorithm in the context of differential machine learning, where differential labels (gradients of labels wrt inputs) are available too. Identify a small number of edgemost examples in the training set, e.g. by Gaussian likelihood of inputs. Recall that 'extreme' scenarios mean extreme \emph{inputs} here, irrespective of labels. Train on sample labels and pathwise differentials for interior examples and ground truth labels, including differentials, for edgemost examples. Assign a larger weight to the extreme examples in the cost function to treat them as a soft constraint. The resulting approximation (provided linear asymptotics) will have the correct asymptotic behaviour beyond edge points, where intercepts and slopes are given by construction by ground truth values and gradients. \figref{edges} displays 1024 inputs sampled from a bidimensional Gaussian distribution, where 16 edge examples are identified by Gaussian likelihood.

\begin{figure}[htp]
\centering
\includegraphics[scale=0.6]{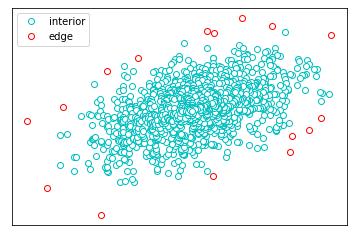}
\caption{Interior and edge training inputs sampled from a bidimensional correlated Gaussian distribution}
\label{fig:edges}
\end{figure}

\subsubsection{Compute ground truth labels with \emph{one} Monte-Carlo path}

Hence, we can control asymptotics effectively with ground truth labels for a small number of edge examples. Contrarily to the simple idea of oversampling extreme scenarios, this method fixes asymptotics without damaging approximation, since it doesn't modify interior examples in any way. The training set is still generated in reasonable time, since only a small number of extreme inputs gets costly ground truth labels. 

However, computation cost may still be considerable compared to a standard LSM dataset. For example, say we simulate 8192 training examples, mostly with sample labels, but in the 64 least likely scenarios, we produce ground truth labels with 32768 nested Monte-Carlo paths. We simulate a total of $64 \times 32768 + 8128 = 2105280$ paths against $8192$, a computation load increased by a factor $257$. 

It turns out that, at least in the context of financial pricing approximation from a Monte-Carlo dataset, we can actually compute ground the true edge values and risks for the cost of \emph{one} Monte-Carlo path, and effectively fix asymptotics without additional cost. 

The common assumption of linear asymptotics comes from that the value of many financial Derivatives converges to \emph{intrinsic value} in extreme scenarios. In general terms, the intrinsic value corresponds to the payoff evaluated on the \emph{forward scenario}, where all underlying instruments fix on their forward values, conditional to initial (extreme) state. See e.g. chapter 4 of Modern Computation Finance (Wiley, 2018) for details.
 
Hence, under the assumption of linear asymptotics, the value and Greeks in edge states are computed for the cost of one Monte-Carlo path, generated from the forward underlying asset prices computed from the (extreme) state variables. By hypothesis of linear asymptotics (a.k.a. asymptotically intrinsic values), this procedure computes correct prices (and Greeks) in extreme scenarios (and only in those). The practical implementation is dependent on the specifics of simulation systems. The idea is illustrated on \figref{edgesLegend}.

\begin{figure}[htp]
\centering
\includegraphics[scale=0.6]{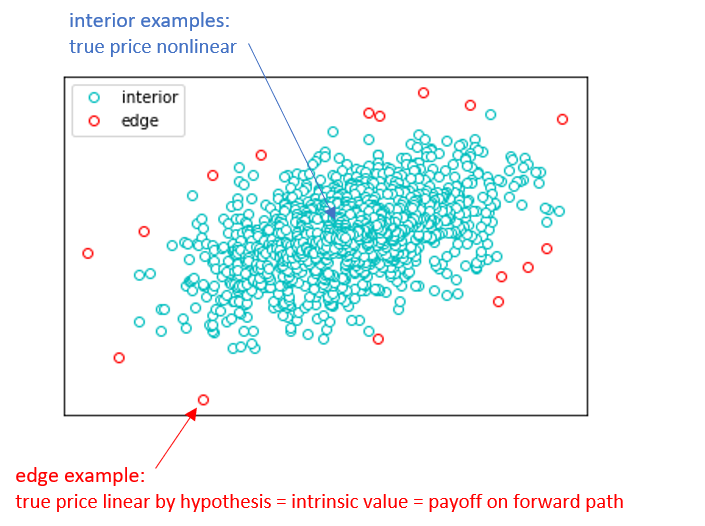}
\caption{Computation of true prices in interior and edge examples}
\label{fig:edgesLegend}
\end{figure}

All labels are computed by sampling payoffs on one path conditional to the scenario, but for edge scenarios, we use a special inner simulation, the 'forward path', where payoffs coincide with intrinsic values and pathwise differentials coincide with true Greeks. With a soft constraint to match these amounts in the cost function, we effectively control asymptotics without additional computational cost or damage to approximation quality.

\section*{Conclusion}

Contrarily to common wisdom, training neural networks does provide some guarantees. In particular, the regression of the output layer on the basis functions identified in the regression layer is guaranteed optimal. We built on this observation to combine classic regression with deep learning in a wide and deep architecture \emph{a la} Google and establish worst case training guarantees. We also discussed the particular effectiveness of the wide and deep architecture when combined with differential PCA presented in \ref{app2}. 

We also covered elementary and advanced asymptotic control algorithms, the most advanced ones, implemented with some effort, being capable of producing correct asymptotics without additional computation cost or stealing data from the interior domain. The algorithm requires differential labels and only works with differential machine learning. In financial Derivatives risk management, differential labels are easily and very efficiently produced with automatic adjoint differentiation (AAD).

\pagebreak 

\bibliography{main}{}
\bibliographystyle{abbrv}

\endgroup
\end{document}